\documentclass[10pt,twocolumn,english,aps,prx,superscriptaddress,showpacs,floatfix]{revtex4}
\usepackage[T1]{fontenc}
\usepackage[latin9]{inputenc}
\usepackage{babel}
\usepackage{amssymb}
\usepackage{graphicx}
\usepackage{esint}
\usepackage[unicode=true,pdfusetitle,
 bookmarks=true,bookmarksnumbered=false,bookmarksopen=false,
 breaklinks=false,pdfborder={0 0 1},backref=false,colorlinks=false]
 {hyperref}


\begin{document}

\title{Boosting Majorana zero modes}

\author{Torsten Karzig}
\affiliation{Department of Physics, California Institute of Technology, Pasadena, California 91125, USA}

\author{Gil Refael}
\affiliation{Department of Physics, California Institute of Technology, Pasadena, California 91125, USA}
\affiliation{\mbox{Dahlem Center for Complex Quantum Systems and Fachbereich Physik, Freie Universit\"at Berlin, 14195 Berlin, Germany}}

\author{Felix von Oppen}
\affiliation{\mbox{Dahlem Center for Complex Quantum Systems and Fachbereich Physik, Freie Universit\"at Berlin, 14195 Berlin, Germany}}
\affiliation{Department of Physics, California Institute of Technology, Pasadena, California 91125, USA}

\begin{abstract}
One-dimensional topological superconductors are known to host Majorana zero modes at domain walls terminating the topological phase. Their nonabelian nature allows for processing quantum information by braiding operations which are insensitive to local perturbations, making Majorana zero modes a promising platform for topological quantum computation. Motivated by the ultimate goal of executing quantum information processing on a finite timescale, we study domain walls moving at a constant velocity. We exploit an effective Lorentz invariance of the Hamiltonian to obtain an exact solution of the associated quasiparticle spectrum and wave functions for arbitrary velocities. Essential features of the solution have a natural interpretation in terms of the familiar relativistic effects of Lorentz contraction and time dilation. We find that the Majorana zero modes remain stable as long as the domain wall moves at subluminal velocities with respect to the effective speed of light of the system. However, the 
Majorana bound state dissolves into a continuous quasiparticle spectrum once the domain wall propagates at luminal or even superluminal velocities. This relativistic catastrophe implies that there is an upper limit for possible braiding frequencies even in a perfectly clean system with an arbitrarily large topological gap. We also exploit our exact solution to consider domain walls moving past static impurities present in the system.
\end{abstract}
\maketitle

\section{Introduction}

Originally, Majorana fermions describe fermionic excitations in relativistic quantum field theories which are their own antiparticles \cite{majorana}. More recently, Majorana bound states or zero-energy Majorana fermions (often also loosely referred to as Majorana fermions) have become a popular and rapidly developing research field in quantum condensed matter physics \cite{alicea_new_2012,beenakker2012}. An important impetus for this field is provided by topological quantum information processing which, in its simplest incarnations, might be based on Majorana bound states \cite{kitaev,nayak_non-abelian_2008}. While, {\em mutatis mutandis}, Majorana bound states retain the field-theory concept of a fermion which is its own antiparticle, their original relativistic nature typically plays no role. In fact, Majorana bound states are predominantly considered as static and localized excitations and their motion, if considered at all, is usually treated as adiabatic. (For notable exceptions, see Refs.~\onlinecite{cheng_nonadiabatic_2011,tsvelik_riding_2012,degottardi_topological_2011,perfetto_dynamical_2013}.)

Indeed, adiabatic motion of Majorana bound states is underlying their nonabelian braiding statistics \cite{moore_read,greiter,nayak_2n,nayak_non-abelian_2008,ivanov,stern} which is a cornerstone for their use in topological quantum computation \cite{nayak_non-abelian_2008}. While information storage would rely on nonlocal qubits exploiting the $2^N$-fold ground state degeneracy in the presence of $2N$ Majorana bound states, it is envisioned that information processing proceeds via adiabatic braiding operations of the Majorana bound states. Unlike the more familiar abelian cases (bosons, fermions, or anyons) in which the wavefunction is multiplied by a phase factor upon particle interchange, braiding of Majorana fermions causes a unitary rotation of the initial wavefunction in the space of degenerate ground states.

An essential building block of braiding operations consists of moving a Majorana bound state through the system at a constant velocity $v$. While adiabatic motion corresponds to the limit of small $v$, it may be desirable to perform quantum information processing on finite time scales and thus at finite values of the velocity $v$. This motivates us in this paper to consider the motion of Majorana bound states at arbitrary velocities. The centerpiece of our work is an exact solution of this problem for a particular model which exploits an emergent Lorentz invariance of the underlying equations and thus reintroduces aspects of relativity into the dynamics of Majorana bound states (albeit again in a somewhat different manner than in the original relativistic-field-theory context). Several key features of our results can indeed be interpreted in terms of familiar effects of special relativity such as Lorentz contraction and time dilation.

Majorana bound states were initially discovered in the condensed matter context as excitations of correlated electron phases, most notably in
certain fractional quantum Hall states \cite{read_green}. Much of the recent excitement in the field stems from a series of proposals that predict Majorana bound states in hybrid systems made of more conventional materials \cite{fu_superconducting_2008, sau_generic_2010, alicea_majorana_2010, lutchyn_majorana_2010, oreg_helical_2010, cook_majorana_2011}. Significant theoretical and experimental efforts have been expended on effectively one-dimensional systems with strong spin-orbit interactions proximity coupled to conventional superconductors. This can be realized following a seminal proposal of Fu and Kane \cite{fu_superconducting_2008,fu_josephson_2009} which is based on topological insulator edge states or following an alternative route which utilizes semiconductor quantum wires \cite{lutchyn_majorana_2010,oreg_helical_2010}. Indeed, several experiments on quantum wires may have already provided evidence for Majorana bound states \cite{mourik_signatures_2012,das_zero-bias_2012,rokhinson,marcus}. While 
braiding operations are ill-defined in strictly one-dimensional systems, the nonabelian statistics of Majorana bound states does survive in networks of quantum wires \cite{alicea_non-abelian_2011}.

\begin{figure}[t]
\begin{centering}
\includegraphics[scale=0.5]{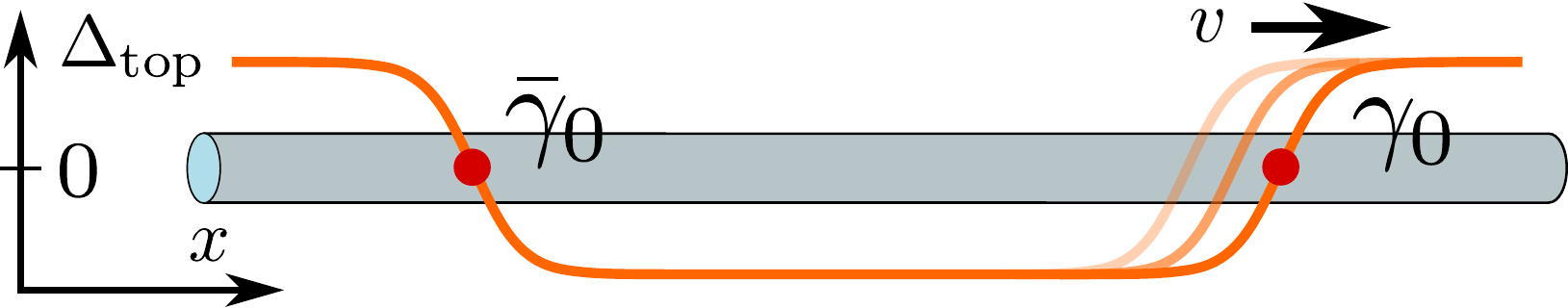}
\par\end{centering}
\caption{Schematic illustration of a moving Majorana bound state $\gamma_{0}$ due to translation of a topological domain wall.\label{fig:Schematic-figure}}
\end{figure} 

In this paper, we consider Majorana bound states moving along such a one-dimensional system. In one-dimensional systems, Majorana bound states (as well as a discrete set of further Andreev bound states) are localized at domain walls between superconducting phases of different topology. These domain walls can be induced in the system by varying parameters -- such as magnetic field, proximity-induced pairing strength, supercurrent, or chemical potential (as controlled by a keyboard of gate electrodes) -- along the wire \cite{lutchyn_majorana_2010,oreg_helical_2010,alicea_non-abelian_2011, romito,van_heck_coulomb-assisted_2012}. By appropriately changing these parameters in time, one can make the domain wall move along the wire at arbitrary velocities $v$ (see Fig.\ \ref{fig:Schematic-figure}). 
We find that the nature of the solution is very different for subluminal ($v<u$) and superluminal ($v>u$) velocities of the domain wall. Here, $u$ denotes the edge-mode velocity of the topological-insulator edge which takes the role of an effective speed of light. In particular, the Majorana bound state exists only for subluminal velocities and dissolves into a continuum of states for luminal or superluminal velocities. While our exact solution is for a particular model of a domain wall, we argue that essential features apply much more generally. Specifically, important aspects carry over not only to more general models of the domain wall but also from topological-insulator edges to quantum-wire-based systems since for parameters in the vicinity of the topological phase transition, the low-energy theory takes on a universal form. 

The paper is organized as follows. In Sec.\ \ref{sec:Model}, we review the model and the phase diagram of a topological superconductor based on a topological-insulator edge proximity coupled to an $s$-wave superconductor and discuss the spectrum and the wavefunctions of Andreev bound states (including the Majorana bound state) for a smooth domain wall. In Sec.\ \ref{sec:Lorentz-invariance}, we establish the effective Lorentz invariance of the Hamiltonian and exploit this invariance to find an exact solution for the problem of a moving domain wall. For subluminal velocities, this solution relies on Lorentz boosting the system to a renormalized static problem, while for superluminal velocities, the solution proceeds via a reference frame in which the problem becomes translationally invariant. We also show that the exact solution has an interesting counterpart in graphene in crossed electric and magnetic fields.  Sec.\ \ref{sec:Physical-consequences} is devoted to a discussion of various physical consequences 
of our exact solution. We discuss the solution for both, subluminal and superluminal velocities as well as the relativistic collapse of the spectrum when reaching the effective speed of light. We also consider the effect of static impurities which, in the comoving frame, act as time-dependent perturbations to the Andreev bound states. In Sec.\ \ref{sec:Generalizations}, we discuss various generalizations of our results, including the generalization from topological-insulator edges to systems based on semiconductor quantum wires. Sec.\ \ref{sec:Conclusions} collects our conclusions. 

\section{Model\label{sec:Model}}

\subsection{Topological-insulator edge}

We begin by reviewing the Hamiltonian of a topological-insulator edge proximity coupled to a superconductor and subject to an applied magnetic field \cite{fu_josephson_2009}. Our starting point is a pairing Hamiltonian in second-quantized form,
\begin{equation}
\hat{H}=\frac{1}{2}\int\mathrm{d}x\hat{\Psi}^{\dagger}(x)\mathcal{H}\hat{\Psi}(x)
\end{equation}
in terms of the Nambu spinors $\hat{\Psi}^{\dagger}(x)=\{\hat{\psi}_{\uparrow}^{\dagger}(x),\hat{\psi}_{\downarrow}^{\dagger}(x),\hat{\psi}_{\downarrow}(x),-\hat{\psi}_{\uparrow}(x)\}$. Here, we indicate operators acting on the many-body Hilbert space with a hat. The resulting Bogoliubov-de Gennes (BdG) Hamiltonian $\mathcal{H}(x)$ takes the form 
\begin{equation}
\mathcal{H}  =  up\sigma_{z}\tau_{z}-\mu(x)\tau_{z}+B(x)\sigma_{x}+\Delta(x)\tau_{x}\,,\label{eq:original_H}
\end{equation}
where $\sigma_{i}(\tau_{i})$ are Pauli matrices acting on spin (particle-hole) space, while $u$ denotes the edge-mode velocity and $\mu,\Delta$, and $B$ are the chemical potential, the superconducting pairing strength, and the magnetic field, respectively. Throughout this paper we set $\hbar=1$.

A competition between $\Delta$ and $B$ leads to two topologically distinct phases, depending on whether the gap is dominated by $B$ or $\Delta$. For spatially constant $\mu$, $B$ and $\Delta$, the two topological phases can be described by the sign of the gap function $\Delta_{{\rm top}}=\sqrt{\Delta^{2}+\mu^{2}}-|B|$ such that $\Delta_{{\rm top}}<0$ ($\Delta_{{\rm top}}>0$) for the $B$- ($\Delta$-)dominated phase. A transition between these phases necessarily requires a closing of the gap $\Delta_{{\rm top}}$. In inhomogeneous systems, it is possible to realize regions of distinct topology. The closing of the gap at the domain wall between these regions leads to localized Majorana bound states. 

\subsection{Solution for a static domain wall}

Choosing $\mu=0$ allows for a straightforward decoupling of the four-component Bogoliubov-de Gennes equation \cite{oreg_helical_2010} (see Sec.~\ref{sub:finite_mu} for a more general discussion). By observing that $\mathcal{H}$ commutes with $\sigma_{x}\tau_{x}$, the BdG equation can be separated into two two-component subspaces described by the Hamiltonians 
\begin{equation}
\mathcal{H}_{\mp}=up\sigma_{z}+\left[B(x)\mp\Delta(x)\right]\sigma_{x}\,,\label{eq:Hpm}
\end{equation}
where the ($\mp$) spaces are spanned by the $\sigma_{x}\tau_{x}$ eigenstates $\{(\mp1,0,0,1),(0,\mp1,1,0)\}$, respectively. (We denote the Pauli matrices in the resulting subspaces also by $\sigma_i$.) For definiteness, we assume that $B,\Delta>0$ in the vicinity of the domain wall which we take to be located at $x=0$. By comparing with Eq.~(\ref{eq:Hpm}), we see that with these choices, only the ($-$) subspace has a vanishing gap, while the ($+$) subspace remains essentially unaffected by the domain wall, retaining a finite gap of $B(x)+\Delta(x)\sim2B(0)$ in its vicinity. Then, the relevant gap function which vanishes at $x=0$ takes the form $\Delta_{{\rm top}}(x)=\Delta(x)-B(x)$. We can therefore focus on the low-energy Hamiltonian 
\begin{equation}
\mathcal{H}_{-}=up\sigma_{z}-\Delta_{{\rm top}}(x)\sigma_{x}\,,\label{eq:Hm}
\end{equation}
which has the form of a two-component Dirac Hamiltonian whose mass term changes sign at $x=0$. It is well known \cite{jackiw_solitons_1976}
that this sign change guarantees the presence of a localized zero-energy  Majorana bound state. Unless the domain wall is abrupt, it will in general bind further Andreev bound states in addition to the Majorana state. To capture their behavior, we consider a model in which the topological gap function $\Delta_{{\rm top}}(x)$ changes linearly in space, $\Delta_{{\rm top}}(x)=bx$. 

Since the spectrum of $\mathcal{H}_{-}$ is symmetric about zero energy, it can be obtained by squaring the Hamiltonian \cite{oreg_helical_2010},
\begin{equation}
\mathcal{H}_{-}^{2}=u^{2}p^{2}+b^{2}x^{2}+ub\sigma_{y}\,.\label{eq:harmonic oscillator}
\end{equation}
Eq.~(\ref{eq:harmonic oscillator}) then takes the form of a harmonic oscillator Hamiltonian. The constant shift of $ub\sigma_{y}$ can cancel the zero-point energy and one thus finds the energy spectrum of $\mathcal{H}_{-}$ as 
\begin{equation}
E_{n}={\rm sign}(n)\sqrt{|n|}\omega\,,\label{eq:v0Spectrum}
\end{equation}
with a single zero-energy (Majorana) eigenstate $n=0.$ Here, $n$ is a (possibly negative) integer, $\omega=\sqrt{2ub}$, and we assumed
$u,b>0$. The finite-energy eigenstates are given by
\begin{equation}
\phi_{n}(x)=\frac{1}{2}\left(i+\sigma_{x}\right)\left(\begin{array}{c}
{\rm sign}(n) g_{|n|-1}(x)\\
g_{|n|}(x)
\end{array}\right)\,,\label{eq:psi_n}
\end{equation}
where the $g_{n}(x)$ denote harmonic oscillator eigenfunctions with oscillator length $\xi=\sqrt{u/b}$. The Majorana wavefunction is localized at the domain wall, has a Gaussian form, and can be read off from Eq.~(\ref{eq:psi_n}) by setting $n=0$, $g_{-1}=0$, and including an additional factor of $\sqrt{2}$ for correct normalization. The result takes the form
\begin{equation}
\phi_{0}(x)=\frac{1}{\sqrt{2\xi\sqrt{\pi}}}\exp\left(-x^{2}/2\xi^{2}\right)\left(\begin{array}{c}
1\\
{\rm i}
\end{array}\right)\,.\label{eq:majorana_solution}
\end{equation}
The solutions of the BdG equation define Bogoliubov quasiparticles through $\hat{\gamma}_{n}=\int{\rm d}x\phi_{n}^*(x)\hat{\Psi}(x)$.
Expressing the two-component subspace in the original four-component Nambu space leads to $\{1,0\}\hat{\Psi}^{\dagger} = -(\hat{\psi}_{\uparrow}^{\dagger}+\hat{\psi}_{\uparrow})$ and $\{0,1\}\hat{\Psi}^{\dagger}=-(\hat{\psi}_{\downarrow}^{\dagger}-\hat{\psi}_{\downarrow})$.
With this construction, it indeed follows that the zero-mode solution Eq.~(\ref{eq:majorana_solution}) defines a Majorana operator with $\hat{\gamma}_{0}=\hat{\gamma}_{0}^{\dagger}$. Note that for the linear dependence $\Delta(x)-B(x)=bx$, the original Hamiltonian (\ref{eq:original_H})
exhibits another Majorana solution $\bar{\gamma}_{0}$ at $x=2\Delta/b$. There, the $\left(\pm\right)$ subspaces switch roles relative to $x=0$ such that the $\left(+\right)$ subspace describes $\bar{\gamma}_{0},$ while the $\left(-\right)$ subspace remains gapped. 

\section{Exact solution for a moving domain wall\label{sec:Lorentz-invariance}}

\subsection{Lorentz invariance}

We now consider a domain wall moving along the topological-insulator edge at a constant velocity $v$. This can be described by the Hamiltonian (\ref{eq:Hm}) with a time-dependent gap function 
\begin{equation}
 \Delta_{{\rm top}}(x,t)=\Delta_{{\rm top}}(x-vt).
\end{equation}
To solve this time-dependent problem, we search for a transformation into an appropriate moving frame. The linear dispersion of the BdG Hamiltonian (\ref{eq:Hm}) suggests that this can be achieved by a Lorentz transformation. 

To explicitly establish the behavior of the time-dependent BdG equation  
\begin{equation}
{\rm i}\partial_{t}\psi^{(v)}(x,t)=\mathcal{H}_{-}(x-vt)\psi^{(v)}(x,t) \label{eq:time_dep_BdG}
\end{equation}
under Lorentz transformations, it is useful to multiply by $\sigma_{x}$. This brings Eq.~(\ref{eq:time_dep_BdG}) into a ``covariant'' form which treats space and time coordinates on an equal footing, 
\begin{equation}
\left[{\rm i}\partial_{t}\sigma_{x}+{\rm i}u\partial_{x}(-{\rm i}\sigma_{y})+\Delta_{{\rm top}}(x-vt)\right]\psi^{(v)}(x,t)=0\,.\label{eq:covariantH}
\end{equation}
Here, the matrices \footnote{Note that the Dirac matrices are usually denoted by $\gamma$, here we avoid this notation because we frequently use the symbol $\gamma$ for other quantities throughout the paper.} $\alpha^{0}=\sigma_{x}$ and $\alpha^{1}=-{\rm i}\sigma_{y}$ indeed fulfill the Dirac algebra $\left\{ \alpha^{i},\alpha^{j}\right\} =2g^{ij}$ with the metric tensor $g^{ij}={\rm diag}(1,-1)$. Eq.~(\ref{eq:covariantH}) therefore takes the form of a two-component Dirac equation with an effective speed of light $u$ and a space-time dependent mass term $-\Delta_{{\rm top}}(x-vt)$. 

We first set the mass constant and review the known invariance \cite{sakurai_advanced_1967} of the Dirac equation under a Lorentz boost
\begin{eqnarray}
x' & = & \gamma_{\Lambda}(x-\beta_{\Lambda}ut)\label{eq:Lorentz_x}\\
ut' & = & \gamma_{\Lambda}(ut-\beta_{\Lambda}x)\,,\label{eq:Lorentz_t}
\end{eqnarray}
with the usual relativistic notations $\beta_{\Lambda}=v_{\Lambda}/u$ and $\gamma_{\Lambda}=1/\sqrt{1-\beta_{\Lambda}^{2}}$ for a boost velocity $|v_{\Lambda}|<u$. In the four-vector notation $x^{\mu}=\{ut,x\}$ and $\partial_{\mu}=\{\partial_{ut},\partial_{x}\}$, the Lorentz boost in Eqs.~(\ref{eq:Lorentz_x}) and (\ref{eq:Lorentz_t}) implies that the derivatives transform as $\partial_{\mu}=\Lambda_{\ \mu}^{\nu}\partial_{\nu}^{'}$, where $\Lambda_{\ \mu}^{\nu}=\partial_{\mu}x'^{\nu}$. To make the Dirac equation (\ref{eq:covariantH}) invariant under the Lorentz boost $\Lambda$, we must also transform the spinor as $\psi'=S\psi$. Multiplying
Eq.~(\ref{eq:covariantH}) by $S$ leads to 
\begin{equation}
\left[{\rm i}S\alpha^{\mu}S^{-1}\Lambda_{\ \mu}^{\nu}\partial_{\nu}^{'}+\Delta_{{\rm top}}\right]\psi'=0\,,\label{eq:boosted_dirac}
\end{equation}
where we used that the mass term is proportional to the unit matrix. The boosted Dirac equation (\ref{eq:boosted_dirac}) is identical to the original one for $S\alpha^{\mu}S^{-1}\Lambda_{\ \mu}^{\nu}=\alpha^{\nu}$. This condition determines the transformation $S$ of the Dirac spinor. One can motivate its form by the following analogy. For a Dirac particle at rest, the Pauli spinor transforms with $S_{R}=\exp\left(\sigma_{x} \sigma_{y} \theta/2 \right)$ under rotations in the $x-y$ plane. A Lorentz boost, on the other hand, can be viewed as a hyperbolic rotation in Minkowski spacetime (i.e., the $x-t$ plane) by an angle $\theta_{\Lambda}={\rm artanh}\left(\beta_{\Lambda}\right)$. It is therefore natural to assume that $S=\exp\left(\alpha^{1}\alpha^{0}\theta_{\Lambda}/2\right)$ which can indeed be checked explicitly. In the following, we will be primarily interested in 
\begin{equation}
S^{-1}={\rm e}^{\sigma_{z}\theta_{\Lambda}/2}=\sqrt{\gamma_{\Lambda}}{\rm diag}(\sqrt{1+\beta_{\Lambda}},\sqrt{1-\beta_{\Lambda}}),
\label{S}
\end{equation}
where we assumed $\beta_{\Lambda}>0$ for the second equality. 

A crucial difference between Eq.~(\ref{eq:covariantH}) and the entirely Lorentz-invariant Dirac equation is the space-time dependence of the mass term. Even though the mass is no longer constant, it is still proportional to the unit matrix such that the above arguments remain untouched. The Dirac equation in the boosted frame can then be obtained by applying the Lorentz boost $\Lambda$ to the argument $x-vt$ of the mass term, yielding the boosted Hamiltonian 
\begin{equation}
\mathcal{H}_{-}^{'}=up'\sigma_{z}-\Delta_{{\rm top}}\left[x'\gamma_{\Lambda}\left(1-\beta\beta_{\Lambda}\right)-ut'\gamma_{\Lambda}\left(\beta-\beta_{\Lambda}\right)\right]\sigma_{x}\label{eq:general_h'}
\end{equation}
with $\beta=v/u$. We can now significantly simplify the problem by appropriate choices of $\beta_{\Lambda}$. We note in passing that Tsvelik employed Lorentz invariance in a related manner to discuss Majorana fermions interacting with fast bosonic fields \cite{tsvelik_riding_2012}.

\subsection{Subluminal motion ($\beta<1$) \label{sub:Subluminal-motion}}

When the domain wall moves at a subluminal velocity $v<u$, we can choose a boosted frame which is moving at the same velocity, $v_{\Lambda}=v$. In this comoving frame, the boosted Hamiltonian (\ref{eq:general_h'}) becomes time independent and takes the form of Eq.~(\ref{eq:Hm}) with a renormalized slope $b'=b/\gamma$ of the topological gap. The renormalization of $b'$ can be understood as a consequence of the familiar length contraction of special relativity. In the comoving frame, the domain wall is at rest and one measures its proper length. Consequently, the size of the domain wall in the lab frame is contracted by a factor of $1/\gamma$. Since, however, the form of $\Delta_{{\rm top}}$ is defined by the lab-frame Hamiltonian, the size of the domain wall increases in the comoving frame, thus reducing its slope $b'$ by a factor of $1/\gamma$. 

With the knowledge of $b'$, the solutions of the BdG equation in the comoving frame can be read off from Eqs.~(\ref{eq:v0Spectrum}) and (\ref{eq:psi_n}). The wavefunctions take the form 
\begin{equation}
\psi_{n}^{(v)'}(x',t')=\gamma^{-1/4}\phi_{n}(x'/\sqrt{\gamma})\exp\left(-{\rm i}E_{n}^{(v)'}t'\right),
\label{wf_comoving}
\end{equation}
with the corresponding energy spectrum
\begin{equation}
E_{n}^{(v)'}=E_{n}/\sqrt{\gamma}\,.\label{eq:eprime}
\end{equation}
The factors of $1/\sqrt{\gamma}$ in these equations follow from the $b'$-dependence of the oscillator length $\xi$ and the frequency $\omega$. 

To complete the solution of the original time-dependent Hamiltonian (\ref{eq:covariantH}), we still need to transform from the comoving frame back into the lab frame. This is achieved through the inverse Lorentz boost
\begin{equation}
\psi_{n}^{(v)}(x,t)=S^{-1}\psi_{n}^{(v)'}(x',t')\,.
\end{equation}
We thus obtain the lab-frame solution
\begin{equation}
\psi_{n}^{(v)}(x,t)=\phi_{n}^{(v)}(x-vt)\exp\left(-\mathrm{i}E_{ n}^{(v)}t\right)\,,\label{eq:subluminal_solution}
\end{equation}
for the wavefunctions, where we defined  
\begin{eqnarray}
\phi_{n}^{(v)}(x) & \!\!= & \!\!\gamma^{1/4}\left(\!\!\begin{array}{cc}
\sqrt{1+\beta}\! & 0\\
0 & \!\sqrt{1-\beta}
\end{array}\!\!\right)\phi_{n}\left(\sqrt{\gamma}x\right){\rm e}^{{\rm i}q_{n}x}\ \ \ \ \ \ \label{eq:phi}
\end{eqnarray}
in terms of $q_{n}=E_{n}^{(v)}\beta\gamma^{2}/u$. The renormalized spectrum in the lab frame is given by 
\begin{equation}
E_{n}^{(v)}={\rm sign}(n)\gamma^{-3/2}\sqrt{|n|}\omega\,.\label{eq:vSpectrum}
\end{equation}
This provides an {\em exact} solution for the problem of a domain wall moving at an arbitrary subluminal velocity $v<u$. 

The renormalization of the energy spectrum in Eq.~(\ref{eq:vSpectrum}) can also be understood by analogy with special relativity. We already argued that the length contraction leads to a factor of $\gamma^{-1/2}$ in the energy spectrum of the comoving frame {[}see Eq.~(\ref{eq:eprime}){]}. The additional factor of $\gamma^{-1}$ originates from the time dilation (i.e., suppression of frequencies) when transforming from the comoving frame back into the lab frame.

The wavefunctions (\ref{eq:subluminal_solution}) can now be employed to define (time-independent) Bogoliubov operators (cp.\ App.\ \ref{appendix})
\begin{equation}
  \hat\gamma^{(v)}_n = \int dx [\psi_{n}^{(v)}(x,t)]^* \hat\Psi(x,t)  
  \label{eq:gamma_n}
\end{equation}
in the lab frame. Note that $\hat\gamma^{(v)}_{-n} = [\hat\gamma^{(v)}_{n}]^\dagger$. This relation follows by explicit calculation and reflects that $\psi_{- n}^{(v)}(x,t) = CT \psi_{n}^{(v)}(x,t)$, where $CT$ denotes the product of charge conjugation $C$ and time reversal $T$ (see App.\ \ref{appendix}). Specifically, this implies that even for the moving domain wall, the $n=0$ solution defines a Majorana operator $\hat\gamma_0^{(v)}$ with $\hat\gamma^{(v)}_{0} = [\hat\gamma^{(v)}_{0}]^\dagger$. 

It is interesting to express the original many-body Hamiltonian $\hat H$ (projected to the $(-)$ subspace) in terms of the operators $\hat\gamma^{(v)}_n$. Expanding the Nambu field operators $\hat\Psi(x,t)$ in the $\hat\gamma^{(v)}_n$ [see Eq.\ (\ref{expansion}) in App.\ \ref{appendix}] and using the time-dependent Bogoliubov-de Gennes equation, we have
\begin{equation}
   {\hat H}_-(t) = \frac{1}{2}\sum_{n,n^\prime} \langle \psi_{ n}^{(v)} | i\partial_t | \psi_{n^\prime}^{(v)} \rangle [\hat\gamma^{(v)}_n]^\dagger  
        \hat\gamma^{(v)}_{n^\prime}.
\end{equation}
Here and in the following, we use the explicit time dependence ${\hat H}_-(t)$ to denote many-body operators in the Heisenberg picture. 
In view of the structure of the explicit wavefunctions $\psi_{\pm n}^{(v)}(x,t)$ in Eq.\ (\ref{eq:subluminal_solution}), this can be written as 
\begin{equation}
   {\hat H}_-(t) = \sum_{n>0} E_n^{(v)} [\hat\gamma^{(v)}_n]^\dagger  \hat\gamma^{(v)}_{n} + v \hat P(t),
\end{equation} 
which involves the total momentum operator 
\begin{equation}
  \hat P(t) = \int dx \sum_{\sigma=\uparrow,\downarrow}\hat\Psi_\sigma^\dagger(x,t) p \hat\Psi_\sigma(x,t),
\end{equation}
with $p=-i\partial_x$. 

\subsection{Superluminal motion ($\beta>1$) \label{sub:Superluminal-motion}}

We will now investigate the case of a domain wall moving at a superluminal velocity $v>u$. Although we can no longer boost to a comoving reference frame, superluminal motion is a perfectly physical scenario in the present context. From Eq.~(\ref{eq:general_h'}), it is clear that we cannot find a transformation to a time-independent $\Delta_{{\rm top}}$ with $\beta_{\Lambda}<1$. In the language of special relativity, the space-time dependence of $\Delta_{{\rm top}}$ switched from space-like to time-like. This, however, allows for a different simplification by choosing $\beta_{\Lambda}=1/\beta$. In this reference frame, the Hamiltonian is spatially independent and the topological gap becomes purely time dependent, taking the form $\Delta_{{\rm top}}(-ut'/\bar{\gamma})$ with $\bar{\gamma}=1/\sqrt{\beta^{2}-1}$. The problem becomes translationally invariant and thus, momentum is a good quantum number. We therefore make the plane-wave ansatz  
\begin{equation}
   \psi_{k}^{'}(x',t') = \bar{\phi}_{k}^{'}(t')\exp\left({\rm i}kx'\right).
\end{equation}   
We emphasize that in contrast to the case of subluminal motion, these solutions are no longer localized. In fact, this also leads to extended states $\psi_{k}(x,t)$ in the lab frame which are labeled by the continuous set of momentum quantum numbers $k$ and thus constitute a continuous spectrum. 

\begin{figure}
\begin{centering}
\includegraphics[scale=0.17]{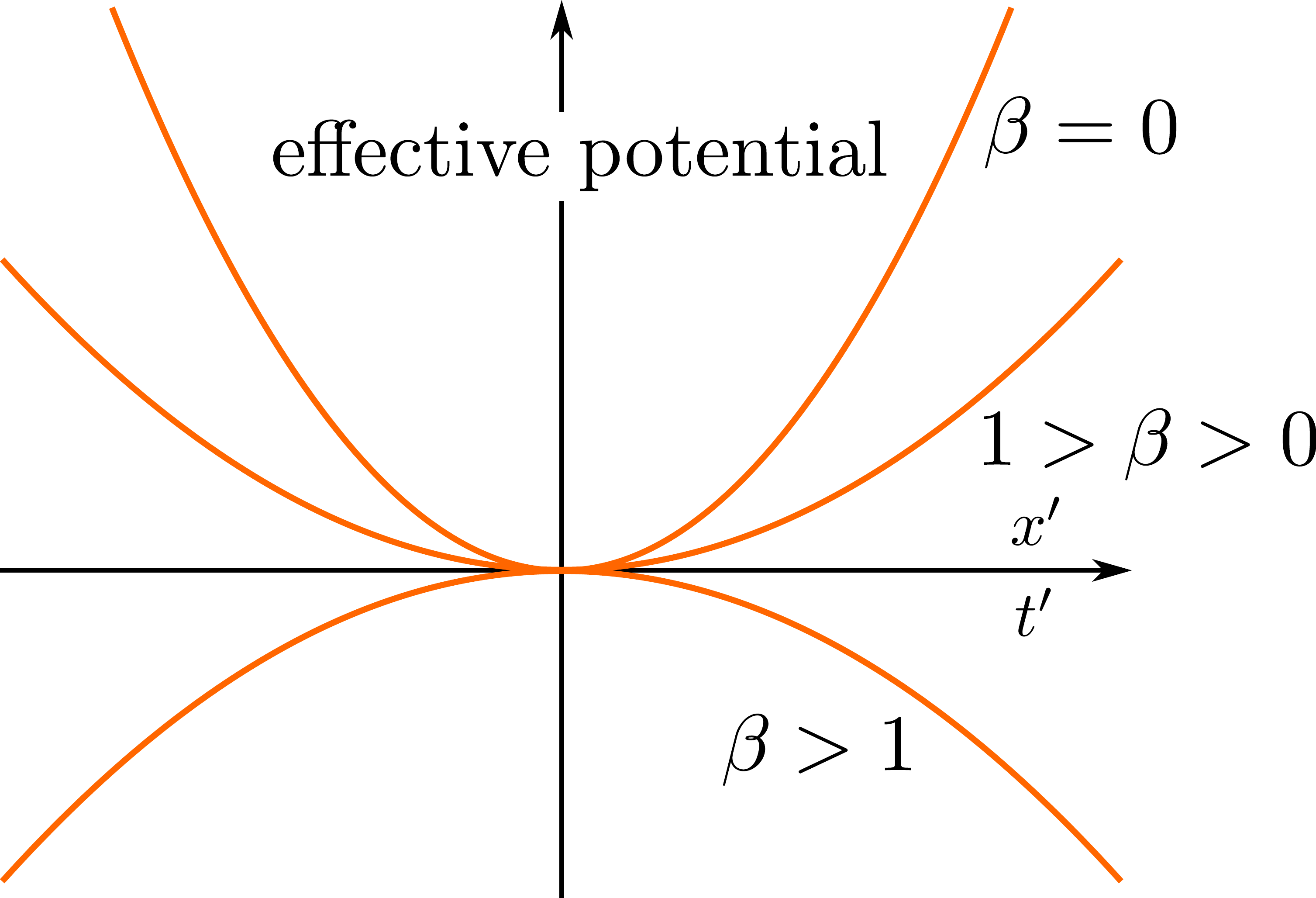}
\par\end{centering}
\caption{Effective potential entering the squared Hamiltonian for the wavefunctions $\phi^{'}(x')$ of the subluminal case ($\beta<1$) and $\bar{\phi}^{'}(t')$ of the superluminal case ($\beta>1$). The transition from a confining to a deconfining potential leads to extended rather than bound states for superluminal motion.\label{fig:Effective-potential}}
\end{figure}

For a linear $\Delta_{{\rm top}}$ and fixed momentum $k$, the time-dependent part $\bar{\phi}_{k}^{'}(t')$ of the wavefunction in the boosted frame is described by the Hamiltonian
\begin{equation}
\mathcal{H}_{-}^{'}(t')=uk\sigma_{z}+\sigma_{x}but'/\bar{\gamma}, \label{eq:LandauZener}
\end{equation}
which equals that of a (rotated) Landau-Zener problem. The solutions $\bar{\phi}_{k}^{'}(t')$ are known to be composed of complex-argument parabolic cylinder functions \cite{zener_non-adiabatic_1932}. To develop intuition for these solutions and to understand the connection to the subluminal case, it is helpful to multiply Eq.~(\ref{eq:LandauZener}) by $\sigma_{z}$. Up to an exchange of time and space variables, the resulting time-dependent
BdG equation
\begin{equation}
\left[{\rm i}\partial_{t'}\sigma_{z}-{\rm i}\sigma_{y}but'/\bar{\gamma}\right]\bar{\phi}_{k}^{'}(t')=uk\bar{\phi}_{k}(t')\label{eq:DGL}
\end{equation}
has the same form as for the static problem Eq.~(\ref{eq:Hm}). The crucial difference is that the mass term is now {\em imaginary}. As a consequence, squaring the Hamiltonian results in a deconfining harmonic potential (see Fig.\ \ref{fig:Effective-potential}) and one
obtains temporally extended states as expected for unitary time evolution.

Specifically, we find that
\begin{eqnarray}
\bar{\phi}{}_{k}^{'}(t') & = & \sum_{\pm}\bigg[\alpha_{\pm}D_{\mp{\rm i}\left(\frac{uk}{\bar{\omega}}\right)^{2}}\left({\rm i}{\rm e}^{\mp{\rm i}\pi/4}\bar{\omega} t'\right)\\
 &  & +\alpha_{\mp}\frac{uk}{\bar{\omega}}{\rm e}^{\mp{\rm i}\pi/4}D_{-1\pm{\rm i}\left(\frac{uk}{\bar{\omega}}\right)^{2}}\left({\rm i}{\rm e}^{\pm{\rm i}\pi/4}\bar{\omega} t'\right)\bigg]\mathbf{e}_{\pm}\nonumber 
\end{eqnarray}
where the $D_{\nu}(z)$ denote parabolic cylinder functions and $\bar{\omega}=\omega/\sqrt{\bar{\gamma}}$. Moreover, $\mathbf{e}_{\pm}=\{1,\pm1\}$ are the eigenvectors of $\sigma_{x}$ and $\alpha_{\pm}$ are free constants to accommodate boundary conditions. The parabolic cylinder functions can be expressed as $D_{\nu}\left(z\right)=2^{-\nu/2}H_{\nu}\left(z\right)\exp\left(-z^{2}/4\right)$ in terms of Hermite functions $H_{\nu}(z)$ and show oscillatory behavior $\sim\exp[\mp{\rm i}(\omega t')^{2}/4\bar{\gamma}]$ as a result of the deconfining harmonic potential (Fig.\ \ref{fig:Effective-potential}).

The solution $\psi_{k}(x,t)$ in the lab frame is again obtained by an inverse Lorentz transformation,  
\begin{equation}
\psi_{k}(x,t)=\bar{S}^{-1}\bar{\phi}_{k}^{'}\left[\bar{\gamma}\left(vt-x\right)\right]\exp\left[{\rm i}k\bar{\gamma}\left(\beta x'-ut\right)\right]\,,
\end{equation}
where $\bar{S}^{-1}=\sqrt{\bar{\gamma}}{\rm diag}\left(\sqrt{\beta+1},\sqrt{\beta-1}\right)$. Of particular interest is the $k=0$ case, which follows by setting $H_{0}(z)=1$ or by directly integrating Eq.~(\ref{eq:DGL}). The corresponding solution reads
\begin{equation}
\psi_{0}\left(x,t\right)=\sum_{\pm}\alpha_{\pm}\exp\left(\mp{\rm i}\bar{\gamma}\left(x-vt\right)^{2}/2\xi^{2}\right)\bar{S}^{-1}\mathbf{e}_{\pm}\,.\label{eq:superluminal_zero}
\end{equation}
This solution is closely related to the subluminal Majorana state which has the form $\propto\exp\left(-\gamma\left(x-vt\right)^{2}/2\xi^{2}\right)$
{[}cp.\ Eqs.~(\ref{eq:subluminal_solution}) and (\ref{eq:majorana_solution}){]}. By lifting the restriction $\beta<1$, we can analytically continue the subluminal solution to the superluminal case via $\gamma\rightarrow-{\rm i}\bar{\gamma}$ and $S^{-1}\rightarrow{\rm diag}(1,{\rm i})\bar{S}^{-1}$. Note that this relation between the sub- and superluminal case is reminiscent of the physics of Cherenkov radiation (see, e.g., Ref.~\cite{cherenkov}) that is emitted by electrons moving faster than the speed of light $\tilde{c}$ of a dielectric medium. There, finite-frequency electromagnetic fields originating from a moving electron change their character from evanescent to propagating waves because the relativistic factor $\sqrt{1-(v/\tilde{c})^2}$, in the same sense as here, turns imaginary when $v>\tilde{c}$. 

Applying the above analytic continuation, the subluminal Majorana solution continues exactly to one of the $k=0$ superluminal solutions in Eq.~(\ref{eq:superluminal_zero}), namely the $(-)$-term. However, in contrast to the subluminal case, Eq.~(\ref{eq:superluminal_zero}) contains a second independent solution. The latter would emerge from an analytic continuation of the unphysical solution $\propto\exp\left(+\gamma \left(x-vt\right)^{2}/2\xi^{2}\right)$. In addition to the continuous nature of the spectrum and the absence of bound states, this is another manifestation of the fact that the topological character is lost in the superluminal case. In fact, this is also confirmed by looking at the Bogoliubov operators associated with the $(\pm)$-terms of Eq.~(\ref{eq:superluminal_zero}), 
\begin{eqnarray}
\hat{\gamma}_{0\pm}^{\dagger} & = & \int{\rm d}x\bigg[-\sqrt{\beta+1}\left(\hat{\gamma}_{\uparrow}^{\dagger}(x)+\hat{\gamma}_{\uparrow}(x)\right)\\
 &  & \mp\sqrt{\beta-1}\left(\hat{\gamma}_{\uparrow}^{\dagger}(x)-\hat{\gamma}_{\uparrow}(x)\right)\bigg]{\rm e}^{\mp{\rm i}\bar{\gamma}\left(x-vt\right)^{2}/2\xi^{2}}.\nonumber 
\end{eqnarray}
These are no longer Majorana operators but instead fulfill $\hat{\gamma}_{0+}^{\dagger}=\hat{\gamma}_{0-}$, suggesting a connection to ordinary Bogoliubov quasiparticles with opposite energies. This can be made explicit for systems where the linear dependence of $\Delta_{{\rm top}}$ saturates at some energy scale $\Delta_{\infty}$. In this case, these solutions do indeed become plane waves asymptotically far from the domain wall with a finite energy $\pm\Delta_{{\rm \infty}}$.

\subsection{Mapping to graphene\label{sub:Mapping-to-graphene}}

Eqs.~(\ref{eq:v0Spectrum}) and (\ref{eq:psi_n}) for the spectrum and wavefunctions of the static domain wall bear a strong resemblance to Landau levels in graphene \cite{goerbig_electronic_2011}. This is not accidental as it is indeed possible to map the static domain-wall Hamiltonian $\mathcal{H}_{-}(x)$ to that of graphene in a magnetic field. In the vicinity of the $\mathbf{K}$ point, the latter takes the form
\begin{equation}
\mathcal{H}_{G}=v_{F}\left(\pi_{x}\sigma_{x}+\pi_{y}\sigma_{y}\right),\label{eq:graphene}
\end{equation}
where $\pi_{i}=p_{i}+eA_{i}$ denotes the kinetic momentum in terms of the vector potential $\mathbf{A}$. The mapping to this Hamiltonian
exploits the fact that up to rescaling, the components of the kinetic momentum operator are canonically conjugate variables, $\left[\pi_{x}, \pi_{y}\right] = -{\rm i}/l_{B}^{2}$.

\begin{figure}
\begin{centering}
\includegraphics[scale=0.12]{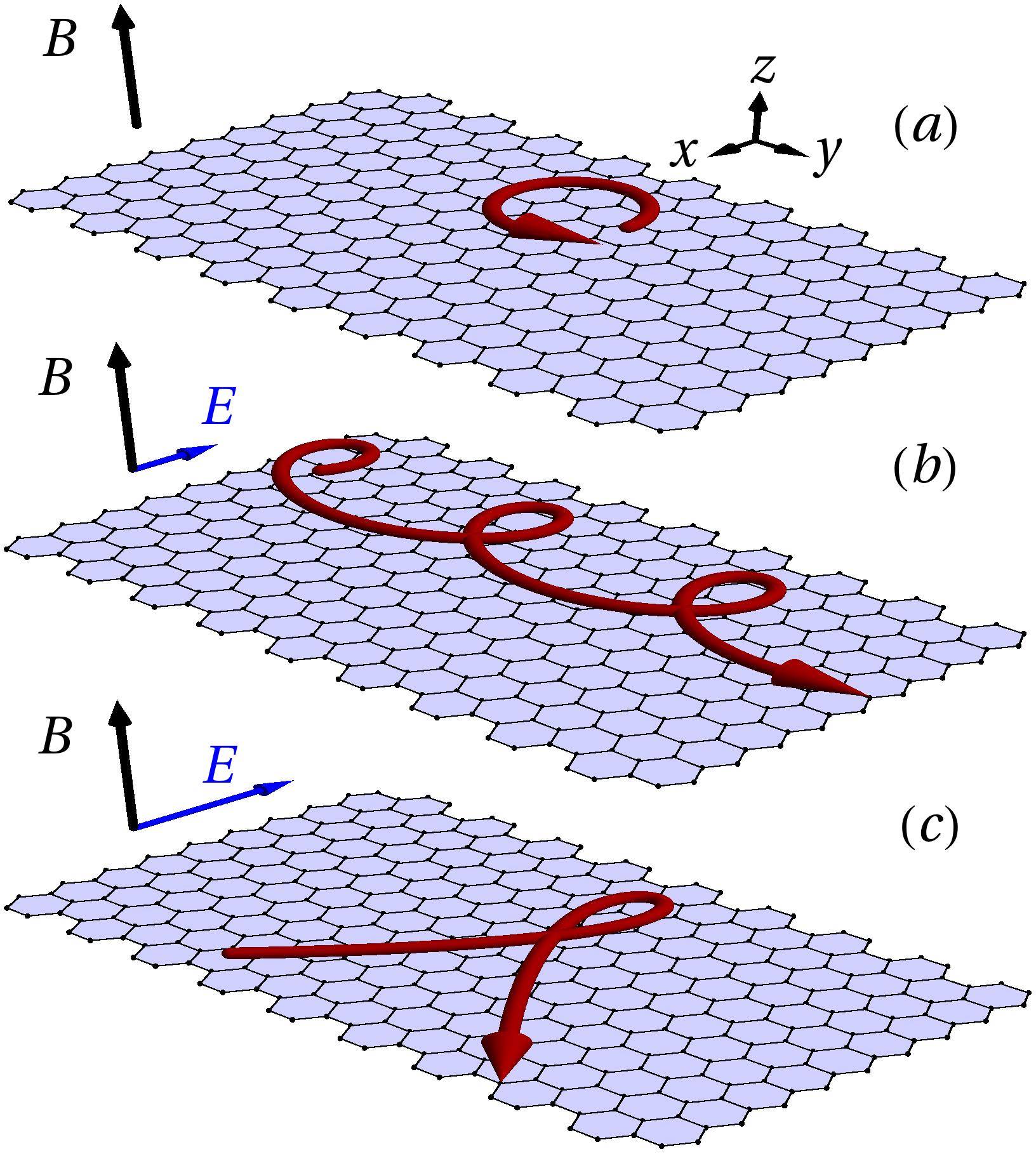}
\par\end{centering}
\caption{Classical trajectories of electrons in graphene subject to crossed magnetic and electric fields. (a) Electric field $\mathbf{E}=0$. 
Electrons perform a circular motion with radius $l_B=1/\sqrt{eB}$. With the mapping developed in the main text, this corresponds to localized bound states of a static domain wall with the Majorana wire pointing in $y$-direction. (b) Small electric fields $|\mathbf{E}|<v_F B$. A drift is superimposed on the closed orbits which are relativistically distorted into ellipses. This maps to the renormalized bound states of the domain wall moving at subluminal velocities. (c) Large electric fields $|\mathbf{E}|>v_F B$. Drift leads to open orbits (hyperbolas in momentum space). This corresponds to the delocalized solutions of a superluminal domain wall. \label{fig:graphene}}
\end{figure}

We start to demonstrate this mapping explicitly by rewriting the domain-wall Hamiltonian of Eq.~(\ref{eq:Hm}) with a static $\Delta_{{\rm top}}(x)=bx$ as
\begin{equation}
\mathcal{H}_{-}=u\left[-(x/\xi^{2})\sigma_{x}+p\sigma_{z}\right]\,.\label{eq:maping}
\end{equation}
Performing a spin rotation about the $x$-axis to rotate $\sigma_{z}$ into $\sigma_{y}$ and renaming $x\rightarrow y$ then brings Eq.~(\ref{eq:maping})
into the same form as the graphene Hamiltonian of Eq.~(\ref{eq:graphene}),
\begin{equation}
\tilde{\mathcal{H}}_{-}=u\left[\left(0-y/\xi^{2}\right)\sigma_{x}+\left(p_{y}+0\right)\sigma_{y}\right].\label{eq:mapping2}
\end{equation}
The static domain-wall problem therefore maps into the $p_{x}=0$ solutions of graphene subject to the vector potential $e\mathbf{A}=\{-y/\xi^{2},0,0\}$ and Fermi velocity $v_{F}=u$. This Landau-gauge vector potential describes a constant magnetic field $eB_{G}=1/l_{B}^{2}$ pointing in $z$-direction such that the oscillator length $\xi=\sqrt{u/b}$ of the localized domain-wall states maps into the magnetic length $l_{B}$ for the graphene Hamiltonian.
To obtain the wavefunctions of the original static domain-wall problem from the corresponding solution for graphene \cite{goerbig_electronic_2011}, one has to undo the spin rotation. This is the origin of the factor $(1-{\rm i}\sigma_{x})/\sqrt{2}$ in Eq.~(\ref{eq:psi_n}).

Interestingly, also the moving domain-wall problem has an analog in graphene. Repeating the same steps as above, the time-dependent domain-wall Hamiltonian can be mapped to graphene subject to a time-dependent vector potential \textbf{$e\mathbf{A}=-\left(y-vt\right)/\xi^{2}\hat{{\rm e}}_{x}$}.
The time dependence of $\mathbf{A}$ implies that in addition to the magnetic field, there is an in-plane electric field $\mathbf{E}=-\partial_{t} \mathbf{A} = -vB_{G}\hat{{\rm e}}_{x}$ pointing in the $x$-direction. The magnitude of the electric field reflects by the velocity $v$ of the domain-wall motion.

The moving domain wall therefore maps to graphene in crossed electric and magnetic fields. The latter problem has been solved in Ref.~\cite{lukose_novel_2007} and the results map exactly to our solutions in Eqs.~(\ref{eq:subluminal_solution}) and (\ref{eq:vSpectrum}).
It is interesting to observe that also the transition from the discrete subluminal to the continuous superluminal spectrum has an analog in graphene. There, the subluminal case corresponds to a magnetic-field-dominated regime with a Landau level spectrum. On the other hand, the superluminal case corresponds to dominating electric fields and metallic transport. In geometric terms, the momentum-space trajectories are closed and ellipsoidal for $v_{F}|\mathbf{B}|>|\mathbf{E}|$ and turn into open hyperbolas once $|\mathbf{E}|>v_{F}|\mathbf{B}|$ \cite{shytov_atomic_2009} (see Fig.~\ref{fig:graphene}).

\section{Physical consequences\label{sec:Physical-consequences}}

\subsection{Renormalization effects}

In Sec.~\ref{sub:Subluminal-motion}, we found that the most immediate effect of subluminal domain-wall motion is a renormalization of the bound states, changing both the spectrum and the spatial extent of the quasiparticle wavefunctions. For a fixed quantum number $n$, the extent of the lab-frame wavefunctions reduces by a relativistic factor of $1/\sqrt{\gamma}$ relative to the static domain wall, cf.\ Eq.~(\ref{eq:phi}). It is interesting to note that this contrasts with the wavefunctions in the comoving frame which actually become more extended by a factor of $\sqrt{\gamma}$, cf.\ Eq.\ (\ref{wf_comoving}). This dichotomy is explained by the fact that this increase is overcompensated by the Lorentz contraction by a factor $\gamma$ when transforming back into the lab frame.

\begin{figure}
\begin{centering}
\includegraphics[scale=0.65]{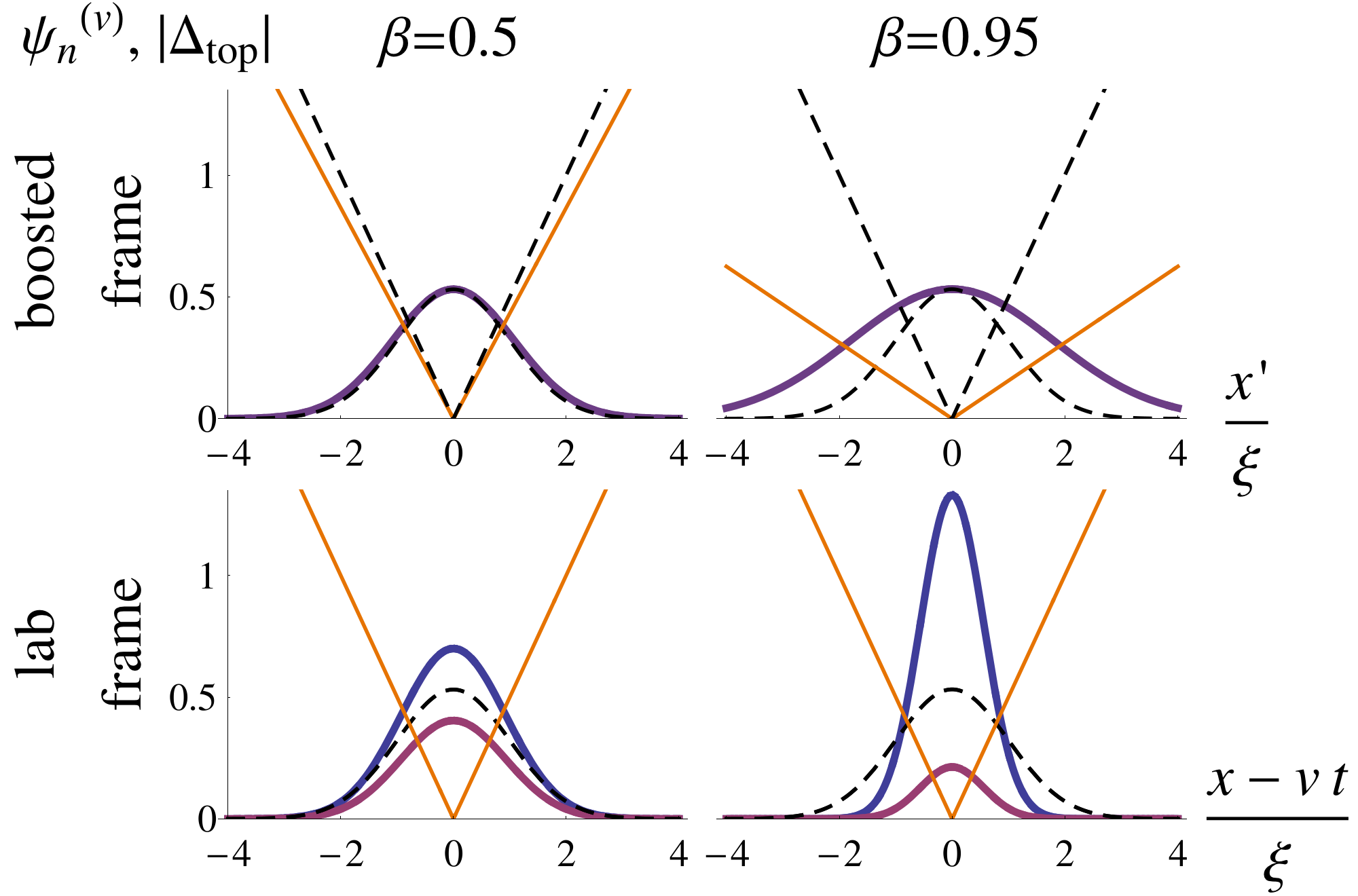}
\par\end{centering}
\caption{Majorana wavefunctions in lab and comoving reference frames, for different velocities as shown in the figure. The solid blue and red graphs show the first and second components of the Majorana wavefunction in the low-energy subspace. The solid orange line plots the absolute value of $|\Delta_{{\rm top}}|\propto|x|$ and thus shows the relativistic renormalization of length scales in the comoving frame. The Majorana wavefunction and the topological gap for a static domain wall are shown for comparison (dashed lines). \label{fig:wavefunctions}}
\end{figure}

As a visualization of these renormalization effects, Fig.~\ref{fig:wavefunctions} shows the form of the Majorana wavefunction at different velocities, in the comoving and lab frame, respectively. Fig.~\ref{fig:wavefunctions} also emphasizes the effect of the spinor rotation $S^{-1}$ when transforming back from the comoving to the lab frame. With the first (second) component of the low energy subspace corresponding to the operators $\hat{\psi}^{\dagger}_{\uparrow(\downarrow)}$ and $\hat{\psi}_{\uparrow(\downarrow)}$, this rotation shifts the weight between spin-up and spin-down particles contributing to the Majorana mode. Since $S^{-1}$ depends linearly on $\beta$ [cf. Eq~(\ref{S})], this shift is the leading consequence of the finite domain wall movement for small velocities.

To estimate the extent of the wavefunction in the lab frame, we note that a state with quantum number $n$ consists of harmonic oscillator wavefunctions with quantum numbers $|n|$ and $|n|-1$. For a static domain wall, the size of the state is thus of order $x_{0}=\sqrt{2|n|+1}\xi$, where $\xi$ denotes the oscillator length. From Eq.~(\ref{eq:phi}), we then find for the moving domain wall that the extent of the wavefunctions is given by $x_{v}=x_{0}/\sqrt{\gamma}$. Interestingly, this implies that the spatial extent of the Majorana bound state is given by $\xi/\sqrt{\gamma}$ which tends towards perfect localization as the domain-wall velocity approaches the effective speed of light, $\beta\rightarrow1$. 

\begin{figure}
\begin{centering}
\includegraphics[scale=0.9]{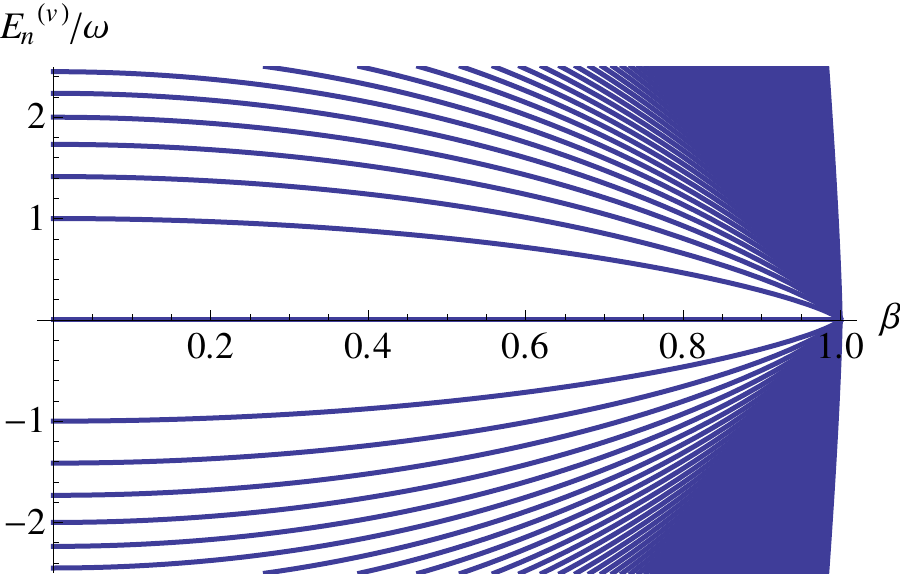}
\par\end{centering}
\caption{Velocity dependence of the bound-state spectrum of a moving domain wall. The spectrum $\sqrt{n}\omega$ of a static domain wall is renormalized by a factor $\gamma^{-3/2}$ for domain walls moving at velocity $v = \beta u$, cf.\ Eq.~(\ref{eq:vSpectrum}). The discrete domain-wall spectrum collapses into a continuous spectrum upon reaching the effective speed of light $u$, where $\beta \rightarrow 1$. \label{fig:spectrum}}
\end{figure}

While the shrinking of the Majorana wavefunction might be viewed as beneficial for the protection of a topological qubit, a finite domain-wall velocity also entails disadvantages which arise from the renormalization of the spectrum. To start with, the level spacings shrink with increasing domain-wall velocity, effectively squeezing the bound-state spectrum (see Fig.~\ref{fig:spectrum}) and making the system more susceptible to perturbations. In addition, we should not only compare the spatial extent of wavefunctions with the same quantum numbers, but also the extent of states with the same energies. In fact, when considering states at a fixed energy $E$, the squeezing of the spectrum implies that we should compare a state with quantum number $n$ for the static domain wall to a state with quantum number $\gamma^{3} n$ of the moving domain wall. When doing so, the spatial extent of the states is larger for the moving domain wall by a factor $\gamma$.

\subsection{Tunneling spectroscopy of the moving bound states}

To further clarify the physical significance of the renormalized spectrum, we note that the  $E_{n}^{(v)}$ can, in principle, be probed by tunneling spectroscopy. Unlike for previous proposals for probing Majorana bound states in tunneling experiments \cite{bolech_observing_2007,law_majorana_2009,flensberg_tunneling_2010,sau_non-abelian_2010}, we consider a tunneling contact that extends over a long distance $L$ along the topological-insulator edge. This can, for instance, be realized by tunneling from a parallel wire. The motivation for this choice is that the bound states of the moving domain wall can be resolved in energy only when they are probed over a sufficiently long time interval \footnote{More specifically, the probing time interval should be longer than the inverse level spacing $1/\omega$ which corresponds to length scales $L\gg \xi$.}.   

Specifically, we consider a tunneling source described by the Hamiltonian
\begin{equation}
 H_S=\sum_{\sigma=\uparrow,\downarrow}\int {\rm d}x\hat{\varphi}^{\dagger}_\sigma(x)\left(\frac{p^{2}}{2m}-    
   \varepsilon_0\right)\hat{\varphi}_\sigma(x)\,,
\end{equation}
where $\varepsilon_0$ accounts for the offset of the band bottom of the source wire relative to the chemical potential $\mu$ of the topological insulator edge which we choose to be $\mu=0$ for simplicity. Tunneling between the source and the domain-wall bound states is described by the tunneling Hamiltonian
\begin{equation}
\hat{H}_T(t)=\eta_0\sum_{\sigma=\uparrow,\downarrow}\int {\rm d}x\hat{\psi}_\sigma^{\dagger}(x,t) \hat{\varphi}_{\sigma}(x,t)+{\rm H.c.\,,}
\end{equation}
with $\eta_0$ measuring the tunneling strength. As we are interested in the low-lying bound states, we project $H_T$ onto the low energy ($-$) subspace by using 
\begin{equation}
[\hat{\psi}_\sigma^{\dagger}(x,t)]_-=-\sum_n [\psi_{n\sigma}^{(v)}(x,t)]^{*}[\hat{\gamma}_n^{(v)}]^{\dagger}\,,
\label{eq:psi_lowenergy}
\end{equation}
which follows directly from the definition of the $(\mp)$ subspaces. Here, $\psi_{n\uparrow(\downarrow)}^{(v)}$ denotes the first (second) component of the solutions of the time-dependent BdG equation $\psi_n^{(v)}$ of Eq.~(\ref{eq:subluminal_solution}). [We also use analogous notation for $\phi^{(v)}_n$.] Combining Eq.~(\ref{eq:psi_lowenergy}) with the expansion of the source field operator in momentum eigenstates,
\begin{equation}
 \hat{\varphi}_\sigma(x,t)=\frac{1}{\sqrt{L}}\sum_k\exp ({\rm i}k x-{\rm i}\varepsilon_{k} t)\hat{c}_k 
\end{equation} 
(in terms of $\varepsilon_k=k^2/2m-\varepsilon_0$), the tunneling Hamiltonian takes the form
\begin{equation}
 \hat{H}_T(t)=\sum_{kn} \eta_{nk}^{\sigma} [\hat{\gamma}_n^{(v)}]^{\dagger}\hat{c}_{k\sigma}{\rm e}^{-{\rm i}\left(\xi_k+eV-vk-E_n^{(v)}\right)t}+{\rm H.c.} .
\label{eq:tunneling}
\end{equation}
Here, we used that $E_n^{(v)}$ is measured relative to the chemical potential $\mu$ and defined $\xi_k=\varepsilon_k-\mu_S$ as well as $eV=\mu_S-\mu$ in terms of the chemical potential $\mu_S$ of the source. Moreover, 
\begin{equation}
 \eta_{nk}^{\sigma}=-\eta_0\int {\rm d} x [\phi_{n\sigma}^{(v)}(x)]^*\frac{1}{\sqrt{L}}{\rm e}^{{\rm i}kx}
\end{equation}
are time-independent tunneling matrix elements. 

The tunneling Hamiltonian in Eq.~(\ref{eq:tunneling}) is effectively that of a more conventional static tunneling problem. If we assume that the initial state at $t=0$ obeys $\langle[\gamma_n^{(v)}]^\dagger \gamma_{n'}^{(v)}\rangle\propto\delta_{nn'}$, the tunneling current can be obtained by a standard calculation. As a result, the energy spectrum $E_n^{(v)}$ manifests itself as (zero-temperature) peaks in the differential tunneling conductance, 
\begin{equation}
 \frac{\partial \langle I\rangle}{\partial V}=\frac{2e^2 L}{v_F} \sum_{n,\pm}\left|\eta_{n,\pm k_F}^\sigma \right|^2\delta(eV\mp vk_F-E_n^{(v)}),
\end{equation}
where we introduced the source Fermi momentum $k_F=\sqrt{2m(eV+\varepsilon_0)}$. Essential features of this result can be understood by physical considerations. First, the strength of the tunneling peaks directly reflects the fact that for the model under consideration, tunneling is momentum conserving. Thus, only the Fourier components $\pm k_F$ of the bound state wavefunctions determine the strengths of the tunneling peaks. Second, the energy shifts $\pm vk_F$ can be thought of as Doppler shifts associated with the relative motion of domain wall and source wire. 

\subsection{Stability of Majorana bound states}

Subluminal motion at a constant velocity can be mapped to a static process and hence does not create excitations or destroy the stability of the Majorana bound state. In particular, consider a system that at $t=0$ starts in a Fock state (e.g., the ground state) $|\Phi\rangle$ with respect to the quasiparticles $\hat{\gamma}_{n}^{(v)}$ as determined by a quasiparticle distribution $f_n$ defined through  $[\hat{\gamma}_n^{(v)}]^\dagger\hat{\gamma}_n^{(v)}|\Phi\rangle=f_n|\Phi\rangle$. This quasiparticle distribution will then stay unchanged
at any later time $t$ with respect to the quasiparticles $\hat{\gamma}_{n,t}^{(v)}=\int {\rm d}x \phi^{(v)}_n(x-vt)\hat{\Psi}(x)$, which are the quasiparticles $\hat{\gamma}_n^{(v)}=\hat{\gamma}_{n,0}^{(v)}$ translated along the system by a distance $vt$. The absence of a motion-induced change of the quasiparticle distribution can be checked explicitly by observing that
\begin{equation}
 [\hat{\gamma}_{n,t}^{(v)}]^\dagger\hat{\gamma}_{n,t}^{(v)}|\Phi(t)\rangle=UU^\dagger [\hat{\gamma}_{n,t}^{(v)}]^\dagger\hat{\gamma}_{n,t}^{(v)} U |\Phi\rangle=f_n|\Phi(t)\rangle\,,
\end{equation}
where we used $U^\dagger\hat{\gamma}_{n,t}^{(v)}U=\gamma_n^{(v)} \exp(-{\rm i} E_n^{(v)}t)$ which follows directly from Eq.~(\ref{eq:gamma_n}).

Even in the absence of acceleration of the domain wall, the Majorana bound states (and hence the associated topological qubits) become unstable when the domain wall moves at superluminal velocities. As shown in Sec.~\ref{sub:Superluminal-motion}, this situation is necessarily described by a time-dependent Hamiltonian and lacks the notion of a localized Majorana mode. This provides a ``speed limit'' for Majorana bound states and thus imposes an upper bound $f_{{\rm max}}$ on the braiding frequency $f_{{\rm b}}$. If the braiding operation requires the domain walls to move along a pathlength $l_{b}$, the maximal braiding frequency is 
\begin{equation}
f_{{\rm max}}=\frac{u}{l_{{\rm b}}}=\frac{\omega}{\sqrt{2}}\left(\frac{\xi}{l_{b}}\right).\label{eq:fmax}
\end{equation}
The length of the braiding path must at least exceed the typical size of the Majorana bound states, $l_{{\rm b}}\gg\xi$, to ensure spatially separated Majorana bound states. The highest possible $f_{{\rm max}}$ is therefore reached in the limit $l_{{\rm b}}\sim\xi$ and of the order of $\omega$, which controls the energy of the first excited bound state level. This coincides with a naive estimate that the braiding frequency should be smaller than the minigap. We emphasize, however, that conventionally, this emerges from an argument about acceleration-induced excitations which is distinctly different from the origin of Eq.~(\ref{eq:fmax}). Moreover, the length of the braiding paths may in general differ in magnitude from the spatial extent of the Majorana bound states due to other design requirements or the need to braid Majorana bound states which are not nearest neighbors. Then, the condition in Eq.\ (\ref{eq:fmax}) is the more stringent one.

The introduced maximal braiding frequency is not only a theoretical upper bound but poses a relevant constraint for experiments. For quantum wires with spin orbit velocity $u=10^4 {\rm m/s}$ (more on this system below) one finds $f_{\rm max}\sim 1 {\rm GHz}$ for braiding path lengths of the order of a micrometer.

\subsection{Impurities\label{sec:Impurities}}

\begin{figure}
\begin{centering}
\includegraphics[scale=0.59]{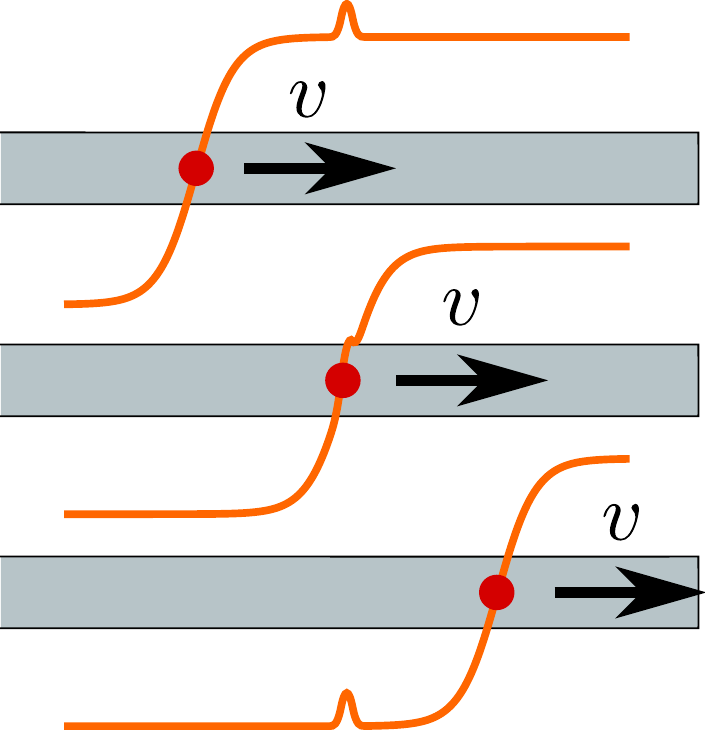}
\par\end{centering}
\caption{In the comoving reference frame, static impurities effectively become time-dependent perturbations for the moving-domain-wall bound states.\label{fig:Static-impurities}}
\end{figure}

So far, we considered a moving domain wall in a clean system. Experimental systems will also contain localized impurities which do not move along with the domain wall. Viewed in the comoving frame of reference, these impurities act as time-dependent perturbations, as illustrated schematically in Fig.\ \ref{fig:Static-impurities}. Thus, these impurities may cause transitions between domain-wall bound states which destroy Majorana-based topological qubits over time. We now apply our exact solution of the moving domain wall for the case of subluminal motion to discuss the transition probabilities 
for such impurity-induced excitations.

Specifically, we consider a general short-range impurity due to local variations of the chemical potential, the magnetic field, or the proximity-induced
superconducting pairing. Within the low-energy subspace, all these impurity types locally modify the magnitude of the topological gap and thus
take the same form 
\begin{equation}
\delta H(x)=\left(-\frac{\Delta}{B}\delta\Delta(x)-\frac{\mu}{B}\delta\mu(x)+\delta B(x)\right)\sigma_{x}.
\end{equation}
(Strictly speaking, the term involving $\delta\mu$ anticipates our discussion of the finite-$\mu$ case in Sec.\ \ref{sec:Generalizations} below.) We assume that the spatial extent of the impurity is small compared to the size of the domain-wall bound states. Then, we can approximate the impurity Hamiltonian as 
\begin{equation}
\delta H(x)=\nu\delta(x)\sigma_{x}.
\end{equation}
where $\nu$ measures the impurity strength. Note that this Hamiltonian is written in the lab frame and that its matrix structure would be modified when transformed into the comoving reference frame. 

We now consider a sufficiently weak impurity for which we can compute the transition amplitude $T_{fi}=\left\langle f|U\left(\infty,-\infty\right)|i\right\rangle$ in Born approximation. Here, $U$ denotes the single-particle (BdG) time-evolution operator. The transition amplitude can be expanded in the Born series
\begin{equation}
T_{fi}=\sum_{n=1}^{\infty}T_{fi}^{(n)},
\end{equation}
where $T_{fi}^{(n)}$ describes the $n$-th order in the Born approximation. Here, we focus on the first-order term
\begin{equation}
T_{fi}^{(1)}=-{\rm i}\int{\rm d}tV_{fi}(t) \label{eq:first order}
\end{equation}
in terms of the transition matrix elements
\begin{equation}
V_{fi}(t) = \int{\rm d}x[\psi_{f}^{(v)}(x,t)]^*\delta H(x)\psi_{i}^{(v)}(x,t).
\end{equation}
Note that the transition amplitudes $T_{fi}^{(1)}$ are Lorentz invariant due to the Lorentz invariance of the volume element $dtdx$ of spacetime. Thus, the transition amplitudes are independent of the reference frame and we could equally well consider the transition amplitudes in the comoving reference frame where $T_{fi}^{(1)}$ describes transitions between stationary eigenstates. 

Evaluating the transition matrix element in the lab frame yields
\begin{eqnarray}
V_{fi}(t) & = & \!\!\frac{\nu\sqrt{\gamma}}{2}\Big[s_{f}g_{|f|-1}(-\sqrt{\gamma}vt)g_{|i|}(-\sqrt{\gamma}vt)\\
 &  & +s_{i}g_{|f|}(-\sqrt{\gamma}vt)g_{|i|-1}(-\sqrt{\gamma}vt)\Big]{\rm e}^{{\rm i}\sqrt{\gamma}\omega_{fi}t},\nonumber 
\end{eqnarray}
where we used the exact wavefunctions (\ref{eq:subluminal_solution}) and defined $\omega_{fi}=E_{f}-E_{i}$ as well as $s_{n}={\rm sign}(n)$. We set $s_{0}=0$ to capture the absence of a $g_{-1}$ term in the Majorana solution. (We leave a factor of $\sqrt{2}$ implicit which would be needed for correct normalization in the $n=0$ case.)

Reflecting the finite interaction time between domain wall and impurity, the transition matrix element $V_{fi}(t)$ is appreciable only during a finite time interval and decays as a Gaussian $\exp(-\gamma v^{2}t^{2}/\xi^{2})$ for large times, i.e., when the domain wall is far from the impurity. The time integral in Eq.~(\ref{eq:first order}) can be performed analytically and is controlled by Franck-Condon-like matrix elements 
\begin{eqnarray}
M_{nn'}(q) & = & \int{\rm d}xg_{n}(x)g_{n'}(x){\rm e}^{-{\rm i}\sqrt{2}qx/\xi}\\
 & = & {\rm e}^{-\frac{1}{2}q^{2}}\sqrt{\frac{m!}{M!}}\left(-{\rm i}q\right)^{M-m}\!\! L_{m}^{M-m}\left(q^{2}\right),
\end{eqnarray}
where $M=\max(n,n')$, $m=\min(n,n')$, and $L_{m}^{M-m}$ denotes the associated Laguerre polynomials. Since the harmonic-oscillator functions are eigenfunction of the Fourier transform, $M_{nn'}(q)$ is essentially an overlap of harmonic oscillator functions which are spatially shifted relative to each other by  $\sqrt{2}q\xi$. In terms of these overlaps, the first-order Born approximation is given by
\begin{equation}
T_{fi}^{(1)}=-\frac{{\rm i} \nu}{2v}\left[s_{f} M_{|f|-1,|i|}\left(\frac{\omega_{fi}}{\beta\omega}\right) +s_{i}M_{|f|,|i|-1}\left(\frac{\omega_{fi}}{\beta\omega}\right)\right]\,.\label{eq:firstorder}
\end{equation}

\begin{figure}[!ht]
\begin{centering}
\includegraphics[scale=0.59]{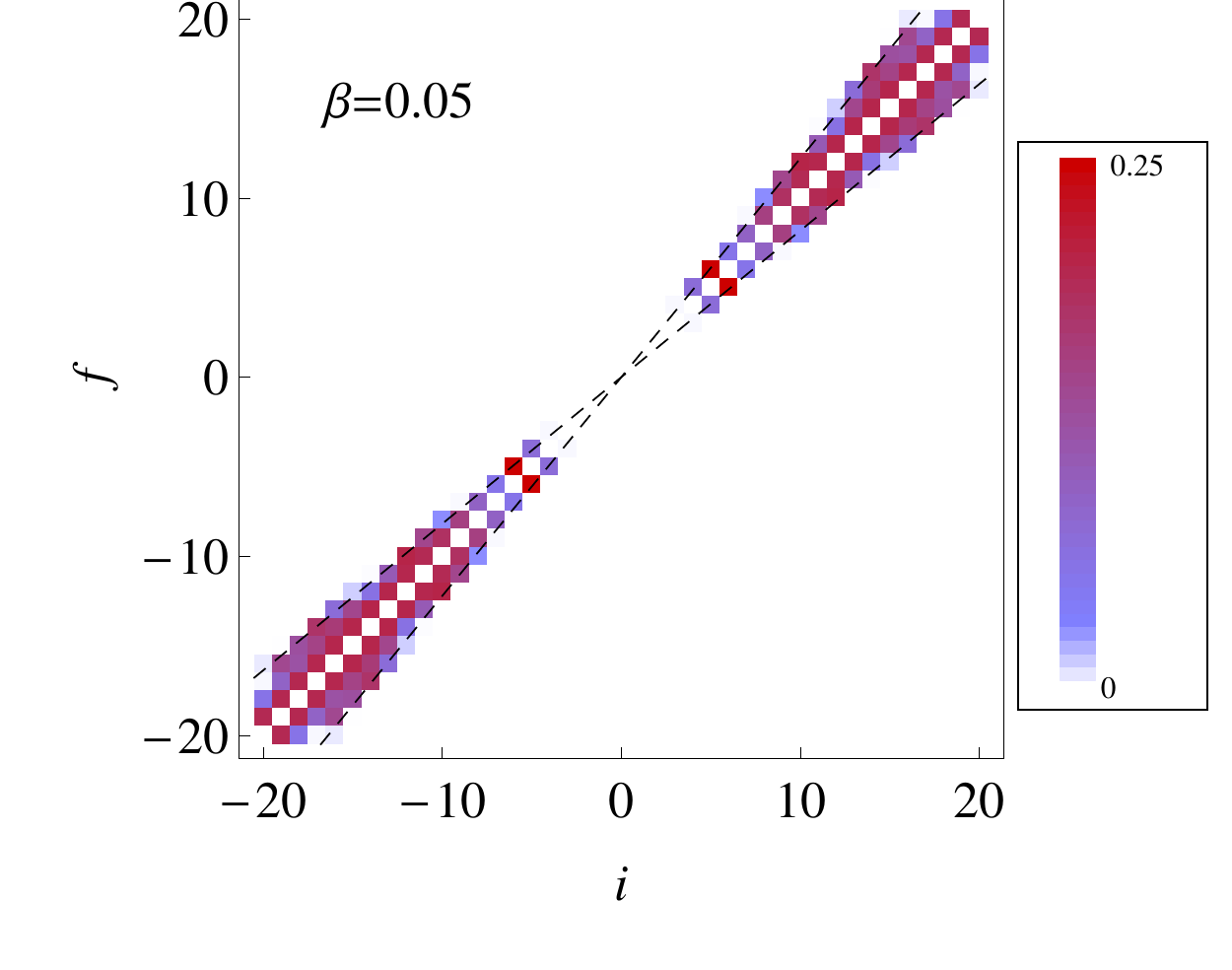}
\includegraphics[scale=0.59]{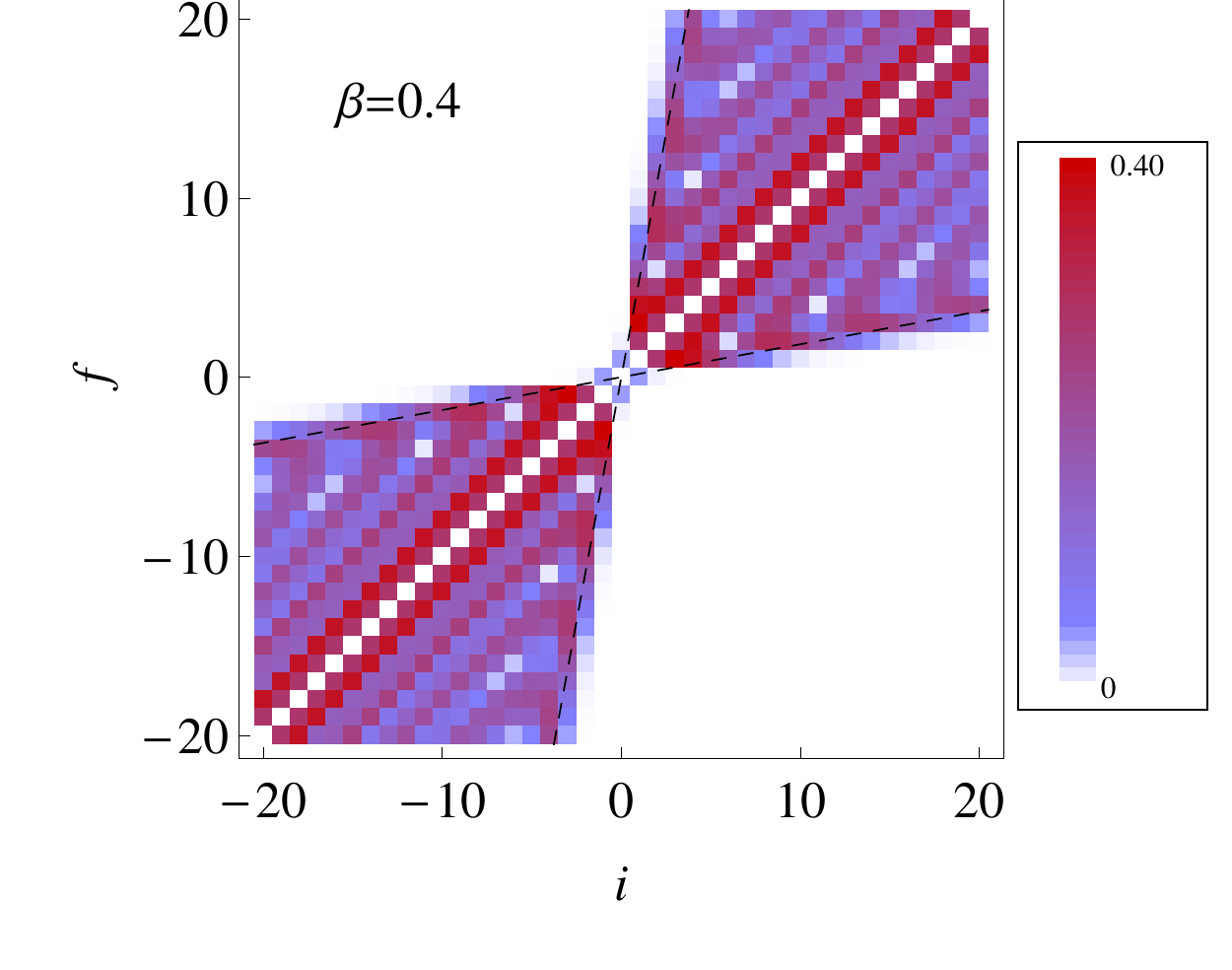}
\includegraphics[scale=0.59]{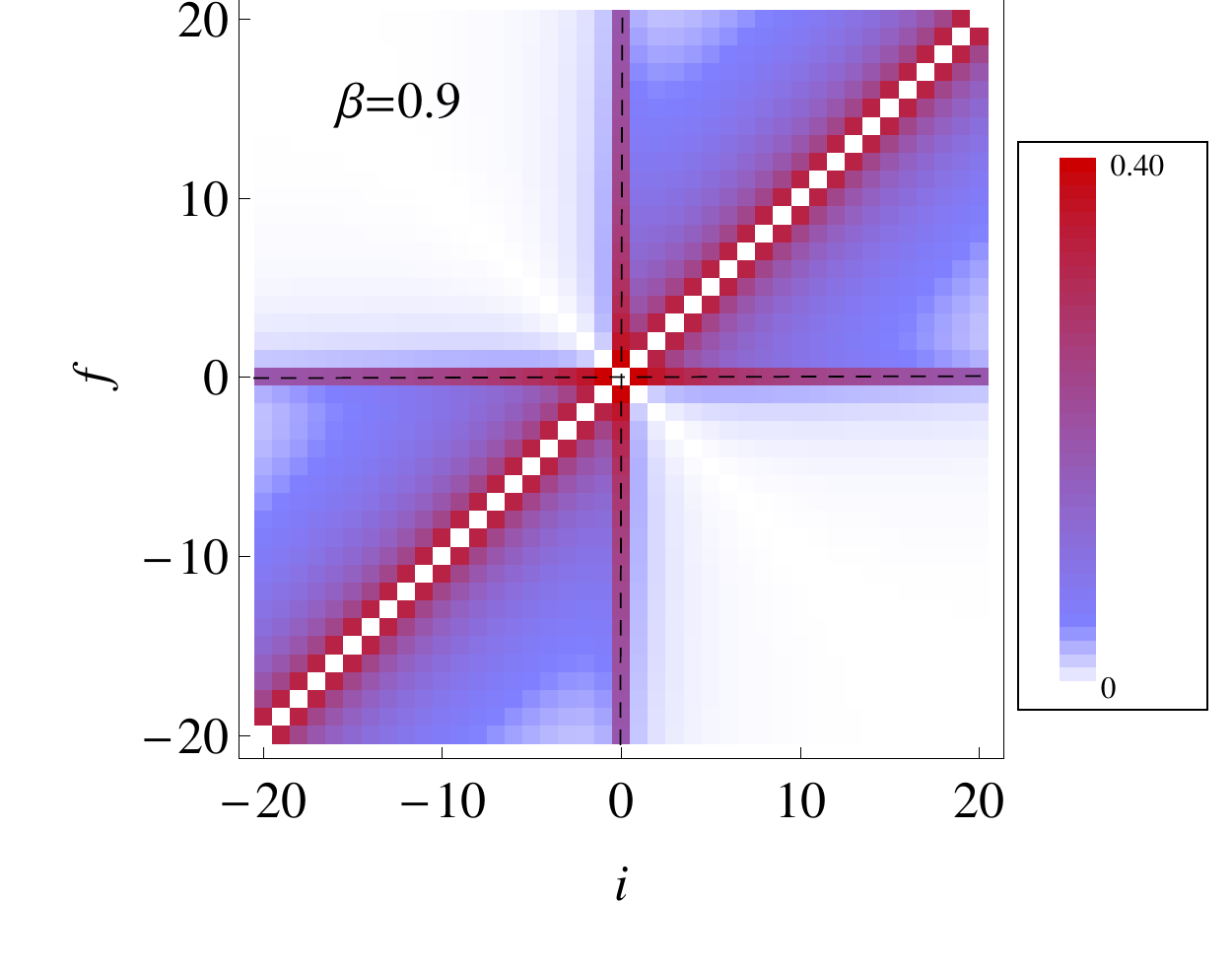}
\par\end{centering}
\caption{Color-scale plots of the impurity-induced transition amplitudes $|T_{fi}^{(1)}|$ in Born approximation [cf.~Eq.~(\ref{eq:firstorder})] as function of initial state $i$ and final state $f$, for velocities $\beta=0.05$, $0.4$, and $0.9$. The dashed lines indicate an estimate for the onset of the Gaussian suppression and are given by $f=(1\pm\beta)^{2}/(1\mp\beta)^{2}i$ (see main text). \label{fig:Visualization}}
\end{figure}

\begin{figure}
\begin{centering}
\includegraphics[scale=0.57]{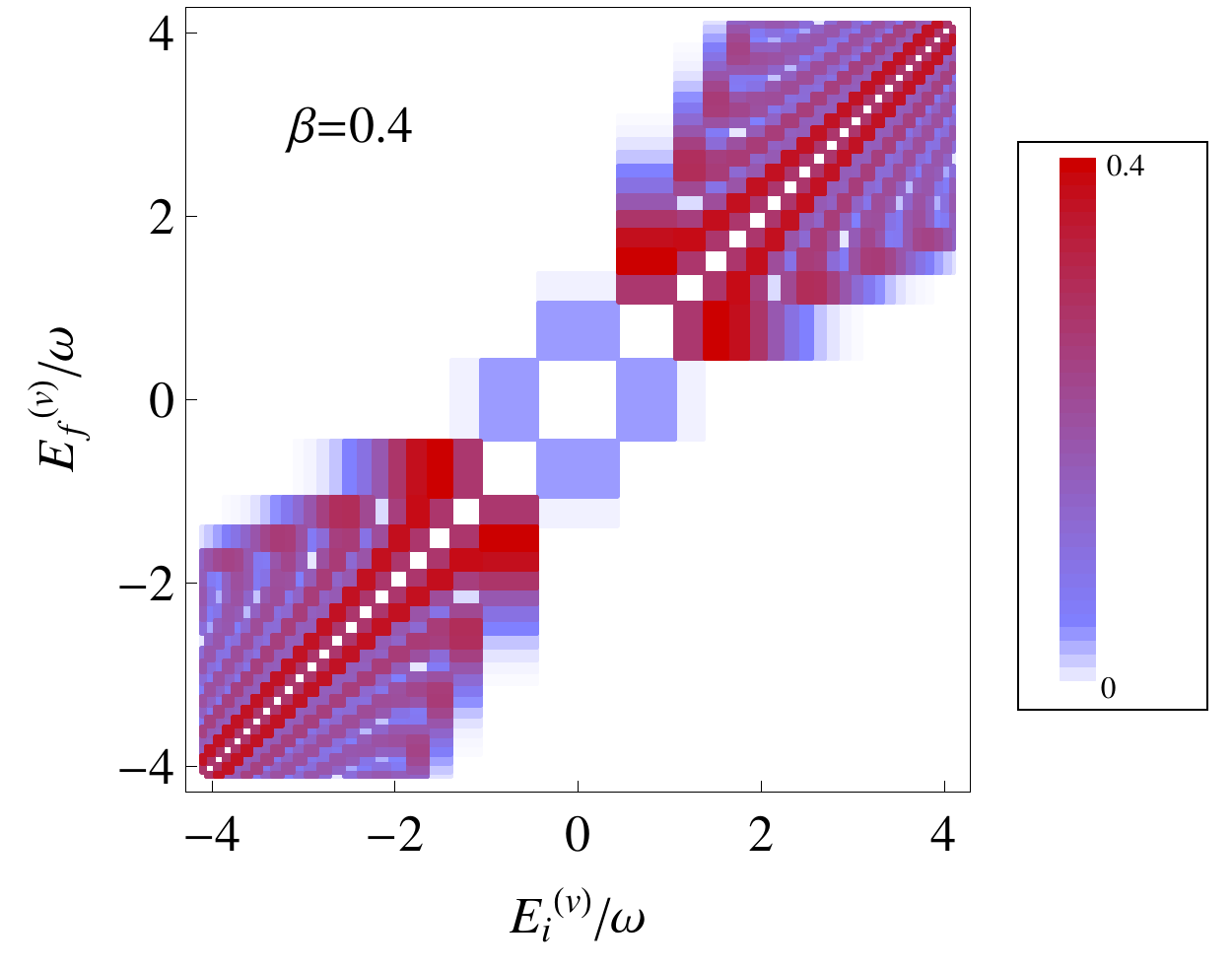}
\includegraphics[scale=0.57]{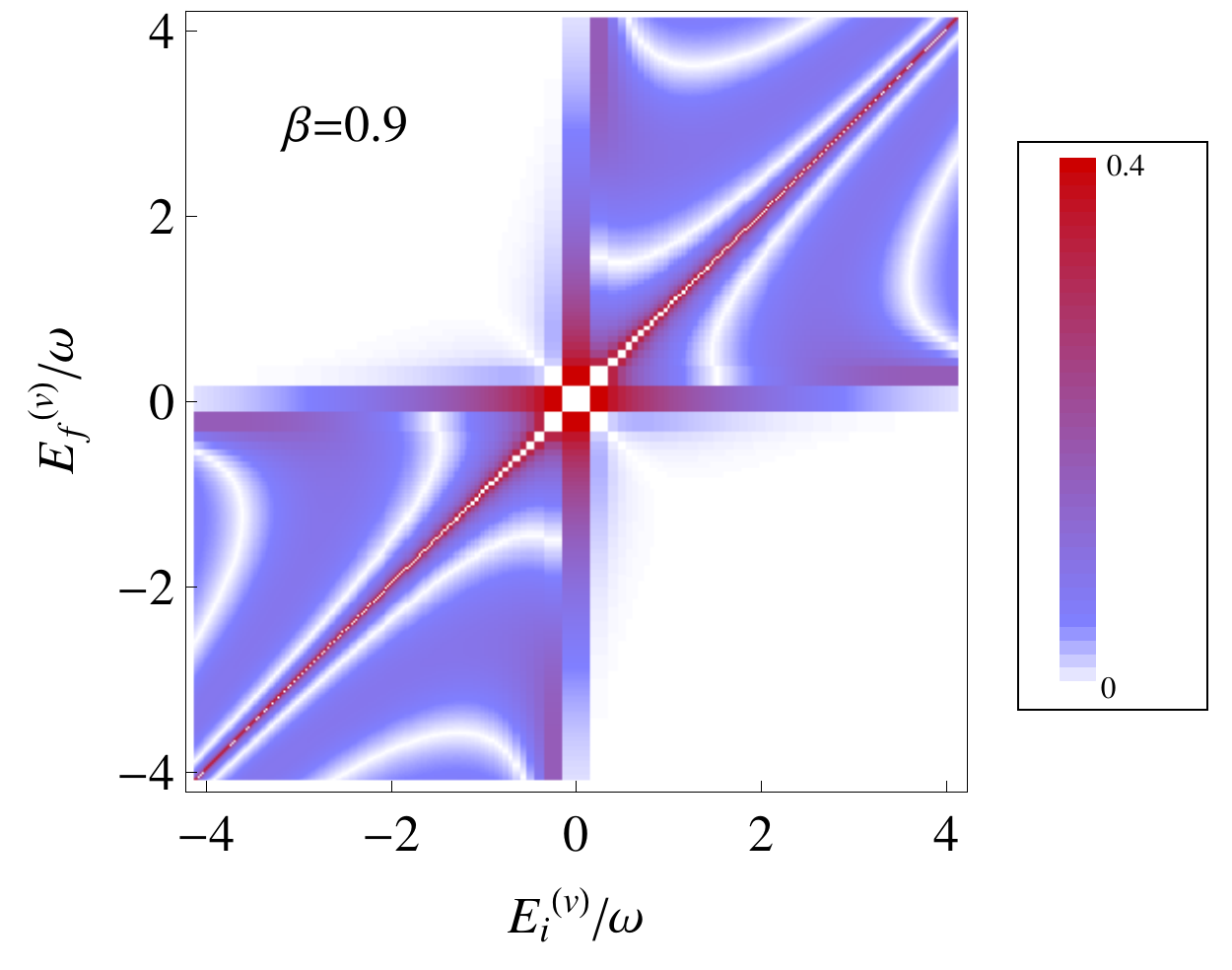}
\par\end{centering}
\caption{Color-scale plots of the transition amplitudes $|T_{fi}^{(1)}|$ for $\beta=0.4$ and $0.9$ as function of the initial and final energies $E_i^{(v)}$ and $E_f^{(v)}$ (instead of the quantum numbers $i$ and $f$ as in Fig.\ \ref{fig:Visualization}). This emphasizes the $\sqrt{n}$ dependence of the energies $E^{(v)}_n$ as well as the squeezing of the energy spectrum for $\beta\to 1$. For $\beta=0.9$, the squeezing implies that there are $\gamma^3\approx12$ times as many states in a fixed energy range compared to the static case.\label{fig:visualization_energy}}
\end{figure}

The transition amplitudes $|T_{fi}^{(1)}|$ are plotted for various domain-wall velocities in Figs.~\ref{fig:Visualization} and \ref{fig:visualization_energy}. The principal characteristics are a strong suppression of transitions between positive and negative energy states as well as a suppression for large energy differences between the initial and final states. These observations can be understood based on the correspondence between the $M_{fi}(q)$ and the overlaps of harmonic-oscillator functions shifted by $\sqrt{2}q\xi$. This shift creates an overall distance between the wavefunction centers given by $\Delta x_{fi}=\sqrt{2} \left(s_{f}\sqrt{|f|}-s_{i}\sqrt{|i|}\right)\xi/\beta$. On the other hand, the sum of the characteristic spatial extents of the wavefunctions is $r_{fi}=\sqrt{2}\left(\sqrt{|f|+1/2}+\sqrt{|i|+1/2}\right)\xi$. Thus, we expect a Gaussian suppression of the overlaps $M_{fi}(q)$ once $|\Delta x_{fi}| > r_{fi}$. Since $\beta<1$, this condition is always fulfilled for $s_{f}s_{i}=-1$, i.e., 
for transitions between positive and negative energy states. This explains the suppression of these transitions as seen in Fig.~\ref{fig:Visualization}. For $s_{f}s_{i}=1$, the condition $\Delta x_{fi}=r_{fi}$ corresponds to $f\approx i(1\pm\beta)^{2}/(1\mp\beta)^{2}$ which defines the limiting straight lines in Fig.~\ref{fig:Visualization} (shown as dashed lines in the figure).

We are particularly interested in the stability of the Majorana bound state, as described by the transition amplitudes with either $i$ or $f$ equal to zero. Note that in addition to their importance for the stability of the Majorana qubit, these amplitudes also describe the dominant excitations from the stationary state in which all negative energy states are occupied and all positive energy states are empty. This state most closely resembles the ground state of the time-independent system and transitions within the sets of positive or negative energy states are forbidden by the Pauli principle. The strong suppression of transitions from negative to positive energy states then makes excitations from or into the zero-energy Majorana mode the dominant process of excitation.

In view of the symmetry of Eq.~(\ref{eq:firstorder}) under the interchange $i\leftrightarrow f$, we can focus on $i=0$ and a positive $f$. As the difference $\Delta x_{f0} - r_{f0}$ increases with increasing $f$, the transition amplitudes fall off exponentially with final-state quantum number $f$. The leading amplitude is thus given by 
\begin{equation}
T_{10}^{(1)}=-{\rm i}\frac{\nu}{\sqrt{2}v}\exp\left[-\frac{1}{2}\left(\frac{1}{\beta}\right)^{2}\right],
\end{equation}
which shows a strong Gaussian suppression for small velocities, $\beta\ll 1$. In fact, in this limit, this suppression is a generic feature of all transition amplitudes since they involve an exponential of the form
\begin{equation}
\label{eq:Tfi}
T_{fi}^{(1)}\sim\frac{\nu}{v}\exp\left[-\left(\sqrt{|f|}-\sqrt{|i|}\right)^{2}/2\beta^{2}\right].\label{eq:first_order_born}
\end{equation}
Matrix elements $T_{f0}$ give the amplitude for creating an excitation $\gamma_f^\dagger\gamma_0=\gamma_f^\dagger(d+d^\dagger)$ and are therefore directly related to a switch in the occupation numer of the corresponding delocalized fermionic zero mode $d$. From Eq.~(\ref{eq:Tfi}) we can conclude that for slow domain-wall velocities impurities are very inefficient in creating excitations, which results in a relatively weak disturbance of Majorana-based topological qubits in this regime.

Interestingly, the first-order Born approximation in Eq.~(\ref{eq:first_order_born}) remains well behaved in the adiabatic limit $\beta\rightarrow0,$ where one would expect the Born approximation to break down. By a saddle point approximation valid for $\beta\ll1$, one can show that this property persists to higher orders of the Born approximation. Indeed, these higher order amplitudes are of the order of
\begin{equation}
T_{fi}^{(n)}\sim\left(\frac{\nu}{v}\right)^{n}\frac{1}{n!}\exp\left[-\left(\sqrt{|f|}-\sqrt{|i|}\right)^{2}/2n\beta^{2}\right].
\end{equation}

\section{Generalizations\label{sec:Generalizations}}

So far we focused on the Hamiltonian (\ref{eq:original_H}) for a topological-insulator edge with a domain wall described by a linearly varying gap function $\Delta_{{\rm top}}$ and zero chemical potential. As we show above, this system is particularly attractive since it allows for an exact analytical solution. At the same time, it is important to understand to which degree this solution also describes the physics of moving domain walls more generally. We will show in the subsequent sections that indeed, essential features of our exact solution carry over to much more general situations.  

\subsection{General domain-wall structures \label{sub:General_gapfunction}}

While a domain wall with a linearly varying gap allows for an exact solution, it is more realistic to consider domain walls for which the gap varies linearly in the vicinity of the domain wall, but eventually saturates to a constant value $\Delta_\infty$ far from the domain wall. In fact, both Figs.\ \ref{fig:Schematic-figure} and \ref{fig:Static-impurities} sketch domain walls with such a structure. 

There are characteristic consequences of the saturation of the topological gap at some distance away from the domain wall, even in the static case. The saturation of the topological gap implies that the domain wall binds only a finite number of discrete states and that the spectrum becomes continuous at energies larger than the maximum topological gap $\Delta_\infty$. Moreover, the bound-state wavefunctions exhibit an exponential decay at large distances from the domain wall and the Gaussian decay of the linear-domain-wall model is limited to intermediate distances. At the same time, unless the domain wall is abrupt on the scale of the oscillator length (as defined by the slope of the topological gap at the domain-wall position), the low-energy spectrum of the static domain wall remains accurately described by the linear-domain-wall model. 
 
The Lorentz transformation of Eq.~(\ref{eq:general_h'}) can be applied to any domain wall which moves at a uniform velocity and is thus described by a topological gap of the form $\Delta_{{\rm top}}(x-vt)$. The mappings to a static problem for $\beta<1$ and to a spatially homogeneous and time-dependent Hamiltonian for $\beta>1$ are possible for any domain-wall structure $\Delta_{{\rm top}}(x-vt)$. This implies that the transition from a low-energy discrete spectrum with a Majorana zero mode for subluminal domain-wall motion to a continuous spectrum for superluminal motion is a generic feature, independent of the domain-wall structure. Thus, the maximal braiding frequency $f_{\rm max}=u/l_b$ in Eq.\ (\ref{eq:fmax}) carries over to this more general case. Moreover, due to the contraction of the bound-state wavefunctions of a moving domain wall, the number of bound states increases with the velocity of the domain wall and diverges when the domain-wall velocity approaches the effective speed of light $u$.

\begin{figure}
\begin{centering}
\includegraphics[scale=0.8]{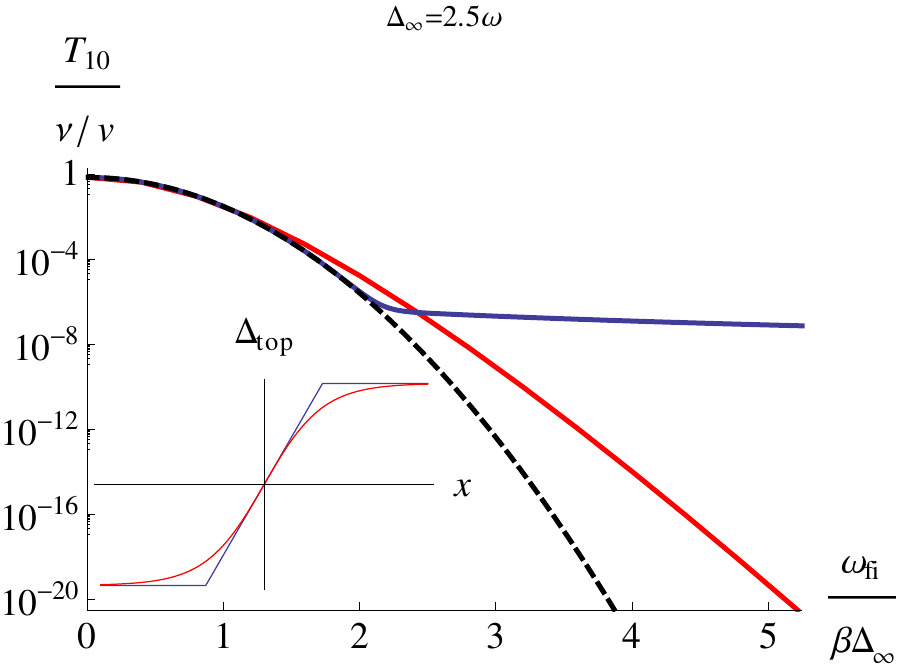}
\par\end{centering}
\caption{$|T_{10}^{(1)}|$ vs $1/\beta$ for different functional forms of $\Delta_{{\rm top}}(x)$ with $\Delta_{\infty}=2.5\omega$. For $\beta>\beta_{\infty}$, the matrix element $T_{fi}^{(1)}$ decays as a Gaussian in $1/\beta$. This Gaussian asymptotics is depicted by the dashed black line. For $\beta<\beta_{\infty}$, the matrix element becomes nonuniversal. As examples, we consider the gap functions shown in the inset. The solid blue line in the main panel corresponds to domain walls with an abrupt switch from linear to constant behavior, cf.\ Eq.~(\ref{eq:saturation}). The red curve shows numerical results for a Fermi-function-like form of $\Delta_{{\rm top}}(x)$.\label{fig:non_universal}}
\end{figure}

The asymptotically exponential decay of the bound-state wavefunctions far from the saturating domain wall also has interesting consequences for the impurity-induced transition rates. Consider a domain wall with a linearly varying topological gap near its center -- with slope $b$ and corresponding frequency $\omega = \sqrt{2ub}$ -- and a saturated gap of $\Delta_{\infty}$ at large distances. Then, the Gaussian time dependence $V_{fi}(t)\sim\exp\left(-\left[\sqrt{\gamma}vt/\xi\right]^{2}\right)$ of the transition matrix element becomes modified into a simple exponential $\exp\left(-\Delta_{\infty}\sqrt{\gamma}vt/u\right)$ for sufficiently large times $\sqrt{\gamma}vt>\Delta_{\infty}/b$. In the first-order Born approximation, the transition amplitude is essentially the Fourier transform of $V_{fi}(t)$ and thus, we find \footnote{The bound results from setting an additional $\cos(\Delta_{\infty}\omega_{fi}/\omega^{2})$ factor in the second term equal to $1$.}
\begin{equation}
 \left|T_{10}^{(1)}\right|\leq\frac{\nu}{\sqrt{2}v}{\rm   e}^{-\frac{1}{2}\left(\frac{\omega_{fi}}{\beta\omega}\right)^{2}}+\beta\sqrt{\frac{2}{\pi}}\frac{\nu}{\omega_{fi}\xi}\frac{\Delta_{\infty}}{\omega_{fi}}{\rm e}^{-2\left(\frac{\Delta_{\infty}}{\omega}\right)^{2}}, \label{eq:saturation}
\end{equation}
when the linear gap saturates abruptly at $\Delta_{\infty}$. In this case, the exponential suppression in $1/\beta$ crosses over to an algebraic one, albeit with a prefactor which is exponentially small in the number of subgap states $(\Delta_{\infty}/\omega)^{2}$. This crossover occurs at a new characteristic velocity $\beta_{\infty}=\omega_{fi}/\Delta_{\infty}$. Interestingly, this implies that an increasing number of subgap states has a \emph{beneficial} effect on the protection of braiding operations against disorder. Note, however, that the behavior for $\beta<\beta_{\infty}$ is nonuniversal as it depends on how $\Delta_{\rm top}(x)$ saturates. When one replaces the abrupt cutoff by a Fermi function profile of the topological gap, an exponential suppression in $1/\beta$ even persists for $\beta<\beta_{\infty}$ (see Fig.~\ref{fig:non_universal}).

\subsection{Nonzero chemical potential\label{sub:finite_mu}}

So far, our considerations focused on the particularly simple case of zero chemical potential, $\mu=0$. For this choice, the Hamiltonian (\ref{eq:original_H}) decouples exactly into two subspaces {[}see Eq.~(\ref{eq:Hpm}){]}. In the vicinity of the domain wall, the low-energy physics is 
determined entirely by one of these subspaces while the other one remains gapped. This decoupling is no longer exact when $\mu$ is nonzero. However, for a saturating domain wall with $\Delta_\infty \ll 2B$, we can still project the problem in a controlled manner to a low-energy subspace by expanding the Hamiltonian about the critical point where the topological gap vanishes. To do so, we start by diagonalizing the original Hamiltonian (\ref{eq:original_H}) for $p=0$ and $\Delta_{\rm top} = \sqrt{\Delta^2 + \mu^2}-B = 0$. Since this puts the system on the phase boundary, we find a low-energy subspace with zero eigenvalue which is spanned by the two eigenvectors $\{(- c_{+},-c_{-},- c_{-},c_{+}),(-c_{-},- c_{+},c_{+},- c_{-})\}$, as well as a high-energy subspace with eigenvalue $2B$ which is spanned by $\{( c_{+},-c_{-}, c_{-},c_{+}),(-c_{-}, c_{+},c_{+}, c_{-})\}$. Here, the eigenvectors are written in terms of $c_{\pm}=[1\pm\Delta(0)/B(0)]^{1/2}$. We can now write $\mu(x) = \mu + \delta\mu(x)$, $\Delta(x) = 
\Delta + \delta\Delta(x)$, and $B(x)  = B + \delta B(x)$, where $\mu$, $\Delta$, and $B$ denote the values at the position $x=0$ of the domain wall, i.e., where the topological gap vanishes. For a saturating domain wall, it is then sufficient to project the Hamiltonian to the low-energy subspace which yields
\begin{equation}
 {\mathcal{H}}_{-}=\tilde{u}p\sigma_{z}-\delta\Delta_{\rm top}\sigma_{x}.
 \label{lowETI}
\end{equation}
This low-energy Hamiltonian is of the same form as for $\mu=0$ but with renormalized parameters which implies that our considerations carry over essentially unchanged to nonzero chemical potentials. Specifically, the topological gap takes the form
\begin{equation}
 \delta \Delta_{\rm top} = \frac{\partial \Delta_{\rm top} }{\partial \mu} \delta \mu +\frac{\partial \Delta_{\rm top} }{\partial \Delta} \delta \Delta +\frac{\partial \Delta_{\rm top} }{\partial B} \delta B 
\end{equation}
and there is a downward renormalization of the mode velocity 
\begin{equation}
\tilde{u}=u\Delta/B=u\sqrt{1-(\mu/B)^{2}}.
\end{equation}
This implies that the effective speed of light of the system is reduced, and with it the critical velocity at which the domain-wall spectrum collapses into a continuum. 

\subsection{Quantum wires with spin-orbit coupling}\label{sec:wires}

Spin-orbit coupling can also induce topological superconductivity in semiconductor quantum wires proximity-coupled to $s$-wave superconductors \cite{lutchyn_majorana_2010,oreg_helical_2010}. The corresponding BdG Hamiltonian takes the form
\begin{equation}
\mathcal{H}  =  [p^2/2m + up\sigma_{z}-\mu(x)]\tau_{z}+B(x)\sigma_{x}+\Delta(x)\tau_{x} . \label{eq:H_wire}
\end{equation}
Our results essentially carry over to this system as well and the reasoning follows the arguments of Sec.\ \ref{sub:finite_mu} for nonzero chemical potential in the topological-insulator model. Indeed, the gap closing at the topological phase transition line of Eq.\ (\ref{eq:H_wire}) always occurs at $p=0$. For this reason, we can expand Eq.\ (\ref{eq:H_wire}) around the critical point by the same sequence of steps as in Sec.\ \ref{sub:finite_mu}. We first identify the low-energy subspace from considering $p=0$ and $\Delta_{\rm top} = \sqrt{\Delta^2 + \mu^2} - B =0$. Thus, the low- and high-energy subspaces are again spanned by the same basis vectors. If the domain wall saturates at a sufficiently small $\Delta_\infty$, we can drop the terms originating from the quadratic dispersion relative to the linear spin-orbit terms. This yields the same low-energy model as in Eq.\ (\ref{lowETI}). Quantitatively, a straight-forward estimate yields the condition $\Delta_{\infty} \ll \varepsilon_{\rm SO}\Delta^2/\mu B$, 
where $\varepsilon_{\rm SO} = mu^2$ measures the strength of the spin-orbit coupling in the wire. Note that this condition is always fulfilled for zero chemical potential.

Finally, we note that nonlinearities in the spectrum can lead to high energy and large momentum states with energies $\varepsilon_k<u k$, where $\varepsilon_k$ is the dispersion of the quantum wire. In the limit $\varepsilon_{\rm SO}\gg B$, this condition applies to the states near $k_F\approx2mu$ where there is a gap of order $\Delta$. These states introduce a new velocity scale $v_h=\varepsilon_{k_F}/k$ and are always irrelevant for the low-energy physics for velocities $v<v_h$. For $v>v_h$, similar to the Landau criterion \cite{landau}, the moving domain wall can in principle mix these large momentum states with the low energy states at $k\approx0$ that contribute to the bound states. For smooth domain walls (on the scale of the spin-orbit length), this mixing will however be exponentially suppressed. Note that in the opposite limit of weak spin-obit coupling $v_h$ is only slightly smaller than the effective speed of light of the system. 

\subsection{Effect of a coupling to bulk modes}

The finite bound state velocity can in principle lead to a mixture of the low energy degrees of freedom and bulk modes (as well as states in the high energy subspace). Similar to the discussion of the Landau criterion in section \ref{sec:wires} this mixture is only effective for sharp enough domain walls. 

The coupling to high energy degrees of freedom becomes most transparent in an alternative description of the time dependent problem, which instead of the Lorentz boost applies the transformation $x'=x-vt$ to the time dependent BdG equation. This results in the effective Hamiltonian
\begin{equation}
 {\mathcal H}_{\rm eff}=\mathcal{H}(x')-vp
\label{eq:Heff}
\end{equation}
acting on ``static`` wavefunctions $\psi(x')$. Bulk states with a static gap $E_{\rm gap}$ and momentum $E_{\rm gap}/v$ have the same energy in the effective Hamiltonian (\ref{eq:Heff}) as those of the low energy subspace. A domain wall of width $w$ can lead to an appreciable mixing of these states when $w \lesssim v/E_{\rm gap}$. For $E_{\rm gap}\gg\omega$ and $w\gg\xi$ the low lying bound states only start mixing with bulk states for velocities $v\gg u$ which is far in the superluminal regime.

\section{Conclusions}
\label{sec:Conclusions}

Every realization of topological quantum computation on a finite time scale has to deal with the dynamic evolution of the corresponding nonabelian quasiparticles. Here, we discussed a crucial ingredient of this dynamics, a moving Majorana-carrying domain wall. Understanding the effects of motion on the Majorana bound state as well as on other components of the spectrum is an essential prerequisite for determining the rate and the space needed to perform Majorana manipulations with high fidelity. 

Alongside these practical motivations of our work, we uncover a set of intriguing connections between the physics of a Majorana bound state moving at a constant velocity, and various seemingly unrelated phenomena. First and foremost, we recapture the `relativistic' nature of the Majorana state. We show that once the Schr\"odinger equation is transformed to the canonical Dirac form, a boost of Majorana states is carried out by means of a standard representation of the Lorentz group, with the mode velocity playing the role of the speed of light. The characteristics of the Majorana bound state exhibit both Lorentz contraction and time dilation, leading to renormalizations of its size and its energy separation from finite-energy states. The ability to apply Lorentz transformations enables us to obtain an exact analytical solution for Majorana-carrying domain walls moving at arbitrary velocities. Second, we find a direct mapping between the problems of a moving domain wall and graphene in crossed electric and magnetic fields. This mapping may help to gain intuition for both systems. 

Our exact solution implies practical bounds on the speed with which Majorana states can be manipulated. Most importantly, Majorana states should not be accelerated to more than the effective speed of light $\tilde{u}$ of the underlying Dirac Hamiltonian. A domain wall moving faster than $\tilde{u}$ leads to an instability similar to superluminal particles in a dielectric that emit Cherenkov radiation. As a result, the discrete domain-wall bound states become continuous and delocalize over the entire system, which leads to a loss of the notion of Majorana bound states. Interestingly, the ensuing upper limit on the braiding frequency is in general more stringent than a naive estimate based on the magnitude of the (mini)gap (cf. Sec. IV C). To push the instability of the spectrum to as large a velocity as possible, one should work close to $\mu=0$. There, $\tilde{u}$ is given by the bare mode velocity, while a finite chemical potential leads to a downward renormalization of $\tilde{u}$ by a factor of $\sqrt{1-(\mu/B)^2}$.

As an application of our exact solution, we study the effect of static short-range impurities which will in general lead to excitations of moving domain-wall bound states and further restrict the applicable velocities. While large velocities $v\sim\tilde{u}$ are clearly detrimental, the Majorana states remain stable for moderate values of $\beta=v/\tilde{u}$. For an ideal linear domain wall, for which the gap increases indefinitely away from its center, excitations of the Majorana bound state are in fact strongly suppressed as a power of $e^{-1/\beta^2}$ (as shown in Sec.\ \ref{sec:Impurities}). For domain walls of a finite size $x_D$, this strong suppression still holds down to velocities $\beta\sim\xi/x_D$ (in terms of the size of the Majorana bound state $\xi$). For smaller velocities $\beta<\xi/x_D$, the suppression becomes nonuniversal, depending on specifics of the domain-wall shape.

As a consequence of the universal low-energy physics underlying the formation of Majorana bound states, the applicability of our theory is rather broad. Specifically, essential aspects of the analysis carry over to various forms of smooth domain walls and (with some restrictions, see Sec.~\ref{sec:wires}) to semiconductor wires as well. (The effective ``speed of light'' is then given by the spin-orbit velocity of the quantum wire).

The analytic solutions presented in the paper are not only of intrinsic interest, but should also be seen as tools that enable further studies of various aspects of Majorana bound state motion. Possible applications vary from dynamical effects on the hybridization of Majorana states and disorder effects on braiding to acceleration effects. In particular, it would be interesting to use the analytic results as a stepping stone for the study of more complex braiding protocols. This opens the search for the best strategies to translate Majorana states while preserving the encoded quantum information.

In addition, it would be interesting to further pursue the connection to special relativity and explore parameter ranges and effects completely inaccessible in other contexts. Perhaps most intriguingly, a constant acceleration of Majorana states may allow for a first observation of the famous Unruh effect \cite{unruh_notes_1976,crispino_unruh_2008}: In relativity, acceleration can be associated with a finite temperature of the system, an effect which is in close correspondence with the Hawking temperature of black holes. In fact, one readily estimates that due to the small effective speed of light in spin-orbit-coupled quantum wires, $u=10^4 {\rm m/s}$, feasible accelerations can lead to experimentally accessible Unruh temperatures. While an experimental detection of the Unruh effect is certainly a difficult task, clever experimental designs might bring it into reach.

\begin{acknowledgments}
It is a pleasure to thank  L.\ Glazman, C.-Y.\ Hou, and D.\ Pesin for helpful discussions. This work was funded by the Institute for Quantum Information and Matter, an NSF Physics Frontiers Center with support of the Gordon and Betty Moore Foundation, the Packard Foundation, a Bessel award of the Alexander-von-Humboldt Foundation, SPP 1285 of the Deutsche Forschungsgemeinschaft, as well as the Helmholtz Virtual Institute ``New states of matter and their excitations.''
\end{acknowledgments}

{\em Note added}.-- During the submission process of this manuscript we became aware of Ref.~\cite{scheurer_nonadiabatic_2013} which studies the non-adiabatic motion of Majorana bound states in quantum wires and shows some overlap to our results.

\appendix

\section{Time-dependent Bogoliubov-de Gennes equation}
\label{appendix}

In this appendix, we briefly review the time-dependent Bogoliubov-de Gennes equations as needed in this paper. Our starting point is the pairing Hamiltonian 
\begin{equation}
\hat{H}=\frac{1}{2}\int\mathrm{d}x\hat{\Psi}^{\dagger}(x)\mathcal{H}\hat{\Psi}(x)
\end{equation}
in terms of the Nambu spinors $\hat{\Psi}^{\dagger}(x)=\{\hat{\psi}_{\uparrow}^{\dagger}(x),\hat{\psi}_{\downarrow}^{\dagger}(x),\hat{\psi}_{\downarrow}(x),-\hat{\psi}_{\uparrow}(x)\}$. We allow for an explicit time dependence of the Bogoliubov-de Gennes Hamiltonian 
\begin{equation}
   {\cal H} =  \left[\begin{array}{cc} h-\mu & \Delta \cr \Delta^* & -(h_T - \mu) \end{array}\right].
\end{equation}
Here, $h$ denotes the normal-state Hamiltonian with its time reverse $h_T = T h T^{-1}$. The time-reversal operator takes the form $T = i\sigma_y K$, where $K$ effects complex conjugation. The Bogoliubov-de Gennes Hamiltonian anticommutes with $CT$, where $C= -i \tau_y$ denotes charge conjugation, 
\begin{equation}
    \{CT,{\cal H}\} = 0.
\end{equation}    
This reflects the fact that the Nambu spinor $\hat\Psi$ satisfies the identity $\Psi^\dagger = (i\tau_y)(-i\sigma_y) \Psi$ due to the doubling of degrees of freedom. In the time-independent case, this implies that for every eigenstate $\psi_n(x)$ of ${\cal H}$ with energy $E$, there is an eigenstate $\psi_{-n}(x) = CT \psi_n(x)$ with energy $-E$. 

The Heisenberg equations of motion $-i\partial_t {\hat \Psi} = [{\hat H},{\hat \Psi}]$ for the Nambu field operator ${\hat \Psi}(x,t)$ take the form
\begin{equation}
   i\partial_t {\hat \Psi} = {\cal H}{\hat \Psi}.
\end{equation}
The ${\hat\Psi}(x,t)$ can be expanded in a set of time-independent fermionic operators $\gamma_n$, 
\begin{equation}
  \hat \Psi(x,t) = \sum_n \psi_n(x,t) \hat\gamma_n = \sum_{n>0} [\psi_n(x,t) \hat\gamma_n + \psi_{-n}(x,t) \hat\gamma_n^\dagger],
  \label{expansion}
\end{equation}
where the wavefunctions $\psi_n(x,t)$ satisfy the time-dependent Bogoliubov-de Gennes equations
\begin{equation}
   i\partial_t \psi_n(xt) = {\cal H}\psi_n(x,t).
\end{equation}
As in the time-independent case, we have divided the complete set of wavefunctions $\psi_n(x,t)$ into two groups (labeled by positive and negative $n$, respectively) which are related by $\psi_{-n}(x,t) = CT \psi_n(x,t)$ and noted that this also implies the relation 
\begin{equation}
  \hat\gamma_{-n} = \hat\gamma_n^\dagger
\end{equation}
for the corresponding field operators 
\begin{equation}
 \hat\gamma_n = \int dx \psi_n^*(x,t) \hat\Psi(x,t).
\end{equation}
The latter satisfy the anticommutation relations $\{\hat\gamma_n,\hat\gamma_{n^\prime}\} = \{\hat\gamma^\dagger_n,\hat\gamma^\dagger_{n^\prime}\} = 0$ and $\{\hat\gamma_n,\hat\gamma_{n^\prime}^\dagger\}=\delta_{n,n^\prime}$ with $n,n^\prime > 0$.  

Throughout most of this paper, we project the original problem described by a $4\times4$ Hamiltonian to a low-energy subspace described by the $2\times2$ Hamiltonian ${\cal H}_-$ in Eq.\ (\ref{eq:Hm}). One readily shows by explicit projection that in the corresponding subspace, the operator $CT$ takes the form
\begin{equation}
   CT  = \sigma_z K.
\end{equation}
With this adjustment, the discussion above also applies to this case.

\bibliographystyle{apsrev4-1_title}
\bibliography{moving_maj}

\end{document}